\documentclass[preprint,amsmath,amssymb,aps]{revtex4}
\usepackage{graphicx}
\usepackage{epsfig}
\usepackage{multirow}
\usepackage{xcolor}
\usepackage{color}
\begin{document}
\title{Semi-classical  features of wobbling and chiral properties in atomic nuclei}

\author{A. A. Raduta$^{a), b)}$, C. M. Raduta $^{a)}$ and R. Poenaru $^{c)}$ }

\affiliation{$^{a)}$ Department of Theoretical Physics, Institute of Physics and
  Nuclear Engineering, 077125 Bucharest, Romania}

\affiliation{$^{b)}$Academy of Romanian Scientists, 3 Ilfov, 050044 Bucharest, Romania}

\affiliation{$^{c)}$ UNESCO cat. II International Center for Advanced Training and Research in Physics (CIFRA), 409, Atomistilor Street, 077125, Bucharest-Magurele, Romania}

\begin{abstract}
A semiclassical approach is used to describe the wobbling and chiral motion in even-even and odd-even nuclei
The trial function involved in the variational equation for the quantal action is a coherent state for the SU(2 ) group associated to a triaxial rotor for the case of  an even-even system and a coherent state for the group SU(2)$\otimes$ SU(2) describing the particle-core system. Application is made for $^{158}$Er and $^{161, 163, 165,167}$Lu, ${135}$Pr.
Th parameters involved in the coherent state expression are complex numbers depending on time and play the role of the phase space generalized conjugate coordinates whose equation of motion may be brought to the Hamilton canonical form.Within a harmonic approximation one analytically obtains the wobbling frequencies which are further used to calculate the excitation energies for the states forming a band. A new procedure to quantize the classical orbitals is proposed. In the new picture the electromagnetic transition  probabilities are calculated both for the in-band and the intraband transitions. Additionally, for odd systems, some other observable like aligned angular momentum, excitation energies relative to a reference spectrum, dynamic moment of energies are calculated and compared with the corresponding experimental data.In the parameter space one identified several nuclear phases with specific properties. An example is given where the wobbling and chiral motions coexist. The existence of a transversal wobbling is widely commented. A new boson representation for the angular momenta is proposed. The approach was successfully applied to $^{135}$Pr. One concludes that semiclassical description is an efficient tool to account quantitatively for the wobbling and chiral properties in nuclei.
\end{abstract}

\maketitle
{\fontsize{11}{11}\selectfont{
\setlength\textwidth{20pc}% %%
\setlength\oddsidemargin{13pt}% %%
\setlength\evensidemargin{13pt}% %%
%\hspace*{-2cm}
%\begin{flushleft}
%\begin{justify}
%\oddsidemargin*{-2cm}

\section{Introduction}
Most of the nuclei from the nuclear chart are axially symmetric and for this reason, the triaxial nuclei were not studied much.
However, the gamma degree of freedom is very important in determining many nuclear properties. This
justifies the attention payed to the $\gamma$ variable even in the early stage of nuclear structure
\cite{Casimir,WilJean,Davy,TerVen}. An extensive study of the triaxial rotor and its coupling with the correlated individual degrees of freedom was achieved in 
Refs.\cite{Toki3,Toki4, Toki5,Tana}.The existence of a $\gamma$ deformed minimum in the potential energy surface leads to specific spectroscopic properties. One of the most exciting features of triaxial nuclei is their possible wobbling motion, which implies a precession of the total angular momentum combined with and oscillation of its projection on the quantization axis around a steady position. The first suggestion for a wobbling motion in nuclei was made by Bohr and Mottelson for high spin states in which the total angular momentum almost aligns to the principal axis with the largest moment of inertia, within the rotor model \cite{BMott}. A fully microscopic description of the wobbling phenomenon was achieved by Marshalek in Ref.\cite{Marsh}. 
Descriptions of the wobbling motion in even-even nuclei are mainly based on classical \cite{Frau01,Frau}, semiclassical \cite{Rad07,Rad017,RaPoAl,PoRa,Bu} and boson expnsion methods \cite{Tana,Badea,Tana2,Oi,Rad017,Rad21}.

Another signature for the $\gamma$ deformed nuclei is the existence of the chiral twin bands which were identifies in system with large transverse magnetic moment. Such a system may consist
of a triaxial core to which a proton prolate  orbital and a  neutron oblate hole orbital are coupled. The maximal transverse dipole moment is achieved when, for example, $\bf{j}_p$ is oriented along the small axis of the core and $\bf{j}_n$ along the long axis and the core rotates around the intermediate axis.   Suppose that the three orthogonal angular momenta form a right trihedral frame. If the Hamiltonian describing the interacting system of protons, neutrons and the triaxial core is invariant to the transformation which changes the orientation of one of the three angular momenta, i.e. the right trihedral frame is transformed to one of a left type, one says that the system exhibits a chiral symmetry. As always happens, such a symmetry is identified when it is broken and consequently to the two trihedrals there correspond distinct energies, otherwise close to each other. Thus, a signature for a chiral symmetry characterizing a triaxial system is the existence of two $\Delta I=1$ bands which are close in energies.  On increasing the total angular momentum, the gradual alignment of $\bf{j}_p$ and $\bf{j}_n$ to the total $\bf{J}$ takes place and a magnetic band is developed. 

The existence of magnetic bands in some nuclei was proved by Frauendorf and Meng \cite{Frau93,Frau97}, being guided by two  results known at that time.  
The first was that of Frisk and Bengtsson \cite{FrBe87} saying that in triaxial nuclei a mean-field cranking solution may exist such that the angular momentum has non-vanishing components on all three principal axes of the inertia ellipsoid. In such case the system under consideration is of chiral type. Indeed, if for example the three angular momentum components form a right-handed reference frame, then their images through a plane mirror form a left-handed frame, which cannot be superimposed on the right-handed one. Therefore, in this case the chiral transformation is supposed to be performed in the space of angular momenta where the time reversal in the position coordinate space becomes a space inversion operation. Obviously, it can be
The second information was provided by the experiment of Petrache {\it et al.} \cite{Pet96} about the existence of a pair of almost degenerate $\Delta I=1$ bands in $^{134}$Pr. This picture is interpreted as being a reflection of a chiral symmetry violation. Some contributions to the field of chiral symmetry were reviewed in Refs.\cite{Rad016}.

In this chapter we shall describe part of our group cotribution to the description of the two signature of the $\gamma$ deformed nuclei.

\section{Wobbling motion in even-even nuclei}
\renewcommand{\theequation}{8.2.\arabic{equation}}
\setcounter{equation}{0}
%\parindent
%\leftflash
%\begin{flushleft}
%\leftsidemargin*{2cm}
%\rightsidemargin*{2cm}
%\begin{flushright}
We suppose that some properties of triaxial nuclei can be quantitatively described by a triaxial rigid rotor.
Therefore, we  consider a  triaxial rigid rotor with the moments of inertia ${\cal I}_k$, k=1,2,3, corresponding to the axes  of the  intrinsic  frame, described by the Hamiltonian:
\begin{equation}
\hat{H}_{R}=\frac{\hat{I}^{2}_{1}}{2{\cal R}_{1}}+\frac{\hat{R}^{2}_{2}}{2{\cal I}_{2}}+\frac{\hat{R}^{2}_{3}}{2{\cal I}_{3}}.
\end{equation}

In principle it is easy to find the eigenvalues of $H_R$ by using a diagonalization procedure within a basis exhibiting the $D_2$ symmetry. However, when we restrict the considerations to the yrast band it is by far more convenient  to use a closed expression for the excitation energies.

We suppose that a certain class of properties of the Hamiltonian $H_R$ can be obtained by solving the time dependent equations provided by the variational principle:

\begin{equation}
\delta \int_{0}^{t} \langle \psi (z)|H-i\frac{\partial}{\partial t^{\prime}}|\psi (z)\rangle dt^{\prime}=0.
\label{varec}
\end{equation}
If the trial function $|\psi (z)\rangle$ spans the whole Hilbert space of the wave functions describing the system, solving the equations provided by the variational principle is equivalent to solving the Schr\"{o}dinger equation associated to $H_R$. Here we restrict the Hilbert space to the subspace spanned by the the variational state:
\begin{equation}
|\psi(z)\rangle ={\cal N}e^{z\hat{R}_-}|IMI\rangle ,
\label{trial}
\end{equation}
where $z$ is a complex number depending on time and $|IMK\rangle $ denotes the eigenstates of the angular momentum operators $\hat{R}^2$, $R_z$ and $\hat{R}_3$ with $R_z$ denoting the angular momentum projection on the OZ axis of the laboratory frame. ${\cal N}$ is a factor which assures that the function $|\psi\rangle$ is normalized to unity: 
\begin{equation}
{\cal N}=(1+|z|^{2})^{-I}.
\end{equation}
$\hat{R}_-$ denotes the lowering operator which for the intrinsic components is :
\begin{equation}
\hat{R}_-=\hat{R}_1+i\hat{R}_2.
\end{equation}
%%%%%%%%%%%%%%%%%%%%%%%%%%%%%%%%%%55
The function $(\ref{trial})$ is a coherent state for the group $SU(2)$ \cite{Kura}, generated by the angular momentum components, and is suitable for the description of the classical features of the rotational degrees of freedom. Due to the supercompletness property, the variational state comprises all basis vectors spanning the Hilbert space. Actually this is the feature which assures
a good approach to the eigenfunctions of $H$. As a matter of fact this will be concretely checked out within the present section.

The averages of $H_R$ and the time derivative operator with the function (2.3), have the expressions:
\begin{eqnarray}
\langle\hat{H}\rangle &=&\frac{I}{4}\left(\frac{1}{{\cal I}_{1}}+\frac{1}{{\cal I}_{2}}\right)+\frac{I^{2}}
{2{\cal I}_{3}}+\frac{I(2I-1)}{2(1+zz^{*})^{2}}\left[\frac{(z+z^{*})^{2}}{2{\cal I}_{1}}-\frac{(z-z^{*})^{2}}{2{\cal I}_{2}}-\frac{2zz^{*
}}{{\cal I}_{3}}\right], \nonumber\\
\langle\frac{\partial}{\partial{t}}\rangle &=&\frac{I(\dot{z}z^{*}-z\dot{z}^{*})}{1+zz^{*}}.
\end{eqnarray}
Denoting the average of $H_R$ by ${\cal H}$, the time dependent variational equation yields:
\begin{equation}
\frac{\partial{\cal{H}}}{\partial{z}}=-\frac{2iI\dot{z}^{*}}{(1+zz^{*})^{2}},\;\;
\frac{\partial{\mathcal{H}}}{\partial{z^{*}}}=\frac{2iI\dot{z}}{(1+zz^{*})^{2}}.
\end{equation}
By a suitable change of variables the classical equations acquire the canonical Hamilton form. Indeed taking the polar form of the complex variable z(=$\rho e^{i\varphi}$) and
\begin{equation}
r=\frac{2I}{1+\rho^2},\;\;0\le r\le 2I.
\end{equation}
the new variables ($\varphi$,r) are canonical conjugate and satisfy the equations:
\begin{equation}
\frac{\partial{\cal{H}}}{\partial{r}}=\dot{\varphi},\;\;
\frac{\partial{\cal{H}}}{\partial{\varphi}}=-\dot{r}.
\label{ecmot}
\end{equation}
Accordingly, $\varphi$ and $r$ play the role of generalized coordinate and momentum respectively.
In the new coordinates, the classical energy function acquires the expression:
\begin{equation}
{\cal H}(r,\varphi)=\frac{I}{4}\left(\frac{1}{{\cal I}_{1}}+\frac{1}{{\cal I}_{2}}\right)+\frac{I^{2}}{2{\cal I}_{3}}+\frac{(2I-1)r(2I-r)}{4I}\left[\frac{\cos^{2}{\varphi}}{{\cal I}_{1}}+\frac{\sin^{2}{\varphi}}{{\cal I}_{2}}-\frac{1}{{\cal I}_{3}}\right].
\end{equation}
%%%%%%%%%%%%%%%%%
Averaging the angular momentum components with the function $|\psi(z)\rangle$ one obtains:
\begin{equation}
\langle I_1\rangle=\frac{2I\rho}{1+\rho^2}\cos\varphi,\;\;\langle I_2\rangle =\frac{2I\rho}{1+\rho^2}\sin\varphi,\;\;\langle I_3\rangle =I\frac{1-\rho^2}{1+\rho^2}.
\end{equation}
Another pair of canonically conjugate coordinates is:
\begin{equation}
\xi=I\frac{1-\rho^2}{1+\rho^2}=\langle I_3\rangle \; \rm{and}\;\phi=-\varphi,
\end{equation} 
Indeed, their equations of motion are:
\begin{equation}
\frac{\partial{\cal H}}{\partial \xi} =-\stackrel{\bullet}{\phi},\;\;
\frac{\partial{\cal H}}{\partial \phi} =\stackrel{\bullet}{\xi}.
\end{equation}
Taking the Poisson bracket defined in terms of the new conjugate coordinates one finds:
\begin{equation}
\{\langle I_1\rangle ,\langle I_2\rangle \}=\langle I_3\rangle, \;\;
\{\langle I_2\rangle ,\langle I_3\rangle \}=\langle I_1\rangle,\;\;
\{\langle I_3\rangle ,\langle I_1\rangle \}=\langle I_2\rangle
\end{equation}
Therefore  the angular momentum components form a classical algebra, $SU(2)_{cl}$, with the inner product $\{ , \}$. The correspondence
\begin{equation}
\{\langle I_k \rangle ,\{ , \}i\}\longrightarrow \{I_k, [ , ]\},
\end{equation}
is an isomorphism of $SU(2)$ algebras, which accomplishes the quantization of classical the angular momentum.

 Suppose we solved the classical equations of motion (\ref{ecmot}) and the classical trajectories given by $\varphi =\varphi(t),\; r=r(t)$ are found.
Due to Eq.(\ref{ecmot}), one finds that the time derivative of ${\cal H}$ is vanishing. This means that the system energy is a constant of motion and, therefore, the trajectory lies on the surface ${\cal H} =const.$. Another restriction for trajectory consists in the fact that the classical angular momentum squared is equal to $I(I+1)$. This restriction is automatically fulfilled by the classical angular momentum. The intersection of the two surfaces, defined by the two constants of motion, 
determines the manifold to which the system trajectory belongs.

Studying the sign of the Hessian associated to ${\cal H}$, one obtains the points where ${\cal H}$ acquires extremal values. Here we consider only the case 
${\cal I}_{1}>{\cal I}_{3}>{\cal I}_{2}$, \, when $(0,I)$ is a minimum point for energy, while 
$(\frac{\pi}{2}, I)$ a maximum.

The second order expansion for ${\mathcal{H}}(r,\varphi)$ around the minimum point, yields:
\begin{equation}
\tilde{\mathcal{H}}(r,\varphi)=\frac{I}{4}\Bigg(\frac{1}{{\cal I}_{2}}+\frac{1}{{\cal I}_{3}}\Bigg)+\frac{I^{2}}{2{\cal I}_{1}}+\frac{2I-1}{4I}\Bigg(\frac{1}{{\cal I}_{3}}-\frac{1}{{\cal I}_{1}}\Bigg)r^{\prime 2}+\frac{(2I-1)I}{4}\Bigg(\frac{1}{{\cal I}_{2}}-
\frac{1}{{\cal I}_{1}}\Bigg)\varphi^{\prime 2}.
\end{equation}
This equation describes an oscillator with the frequency:
\begin{equation}
\omega_I=\left(I-\frac{1}{2}\right)\sqrt{\Bigg(\frac{1}{{\cal I}_{3}}-\frac{1}{{\cal I}_{1}}\Bigg)\Bigg(\frac{1}{{\cal I}_{2}}-\frac{1}
{{\cal I}_{1}}\Bigg)}.
\end{equation}
This frequency is associated to the precession motion of the angular momentum around the OX axis.
In our description, the yrast band energies are, therefore, given by:  
\begin{equation}
E_I=\frac{I}{4}\Bigg(\frac{1}{{\cal I}_{2}}+\frac{1}{{\cal I}_{3}}\Bigg)+\frac{I^{2}}{2{\cal I}_{1}}
+\frac{\omega_I}{2}.
\label{yra}
\end{equation}
%%%%%%%%%%%%%%%%%%%%%%%%%%%%%%%%%%%%%%%%%%%%%%%%%%%%%%%%%%%%%%%%%%%%%%%%%%%%%%%%%
The classical energy function can be further quantized following the following scheme.
One  first expresses ${\cal H}$ in terms of the complex conjugate variable $C_1$ and $B^*_1$
\begin{equation}
{\cal C}_1=\sqrt{2I}\sqrt{\frac{2I-r}{r}} e^{-i\varphi},\;\;{\cal B}^*_1=\frac{1}{\sqrt{2I}}\sqrt{r(2I-r)} e^{i\varphi}.
\end{equation}
and then replacing these by the bosons $b$ and $b^+$ one obtains the Dyson boson representation of $H_R$ denoted hereafter by $H_D$. Although  this boson operator is not Hermitian it has real eigenvalues \cite{Ogu}. We searched for the eigenvalues of $H_D^{\dagger}$ by using the Bargmann representation of the boson operators \cite{Bar,Janc,Jan}:
\begin{equation}
b^{\dagger}\to x,\;\;b\to\frac{d}{dx}
\label{Bquant}
\end{equation}
In this way the eigenvalue equation of $H_D^{\dagger}$ is transformed into a differential equation:

\begin{equation}
\left[\left(-\frac{k}{4I}x^4+x^2-kI\right)\frac{d^2}{dx^2}+(2I-1)\left(\frac{k}{2I}x^3-x\right)\frac{d}{dx}-k\left(I-\frac{1}{2}\right)x^2\right]G=E^{\prime}G.
\label{ecdif}
\end{equation}
where
\begin{equation}
k=\frac{\frac{1}{{\cal I}_1}-\frac{1}{{\cal I}_2}}{\frac{1}{{\cal I}_1}+\frac{1}{{\cal I}_2}-
\frac{2}{{\cal I}_3}}.
\end{equation}
It can be easily proved that this equation can be brought to  the algebraic form of the 
Lam\'{e} equation \cite{Bate,Byer}. 
Performing now the change of function and variable:
\begin{equation}
G=\left(\frac{k}{4I}x^4-x^2+kI\right)^{I/2}F,\;\;\;t=\int_{\sqrt{2I}}^{x}\frac{dy}{\sqrt{\frac{k}{4I}y^4-y^2+kI}},
\label{tfuncx}
\end{equation}
Eq.(\ref{ecdif}) is transformed into a second order differential Schr\"{o}dinger equation:
\begin{equation}
-\frac{d^2F}{dt^2}+V(t)F=E^{\prime}F,
\label{Schr}
\end{equation}
with
\begin{equation}
V(t)=\frac{I(I+1)}{4}\frac{\left(\frac{k}{I}x^3-2x\right)^2}{\frac{k}{4I}x^4-x^2+kI}-k(I+1)x^2+I.
\label{potent}
\end{equation}

The considered ordering for the moments of inertia is such that $k>1$. Under this circumstance the potential $V(t)$ has two minima for $x=\pm\sqrt{2I}$, and a maximum for x=0. 

The minimum value for the potential energy is:
\begin{equation}
V_{min}=-kI(I+1)-I^2.
\end{equation}
Note that the potential is symmetric in the variable x. Due to this feature the potential behavior  around the two minima are identical.
To illustrate the potential behavior around its minima we make the option for the minimum $x= \sqrt{2I}$. To this value of $x$ it corresponds, $t=0$. Expanding $V(t)$ around $t=0$ and truncating the expansion at second order we obtain:
\begin{equation}
V(t)=-kI(I+1)-I^2+2k(k+1)I(I+1)t^2.
\end{equation}
Inserting this expansion in Eq.(\ref{Schr}), one arrives at a Schr\"{o}dinger equation for an oscillator. The eigenvalues are
\begin{equation}
E_n^{\prime}=-kI(I+1)-I^2+\left[2k(k+1)I(I+1)\right]^{1/2}(2n+1).
\end{equation}  
The quantized Hamiltonian associated to ${\cal H}$, i.e. $H^{\dagger}_{D}$, has an eigenvalue which is obtained  from the above expression. The final result is:
\begin{equation}
E_n=\frac{I(I+1)}{2{\cal I}_1}+\hbar\omega_I(n+\frac{1}{2}).
\label{wfor}
\end{equation}
where 
\begin{equation}
\omega_I = \left[\left(\frac{1}{{\cal I}_2}-\frac{1}{{\cal I}_1}\right)\left(\frac{1}{{\cal I}_3}-\frac{1}{{\cal I}_1}\right)I(I+1)\right]^{1/2},
\end{equation}
defines the wobbling frequency of the angular momentum.

The Bargmann representation of the angular momentum components is obtained by inserting the correspondence (\ref{Bquant})
into the Dyson boson expansion. The result is:
\begin{equation}
I_+=\sqrt{2I}x,\;\;I_-=\sqrt{2I}(\frac{d}{dx}-\frac{x}{2I}\frac{d^2}{dx^2}),\;\;I_0=I-x\frac{d}{dx}.
\end{equation}
From these expressions one may derive the angular momentum component $I_1$, which may be further averaged with the wave function provided by the Schrodinger equation for a given value of I. As a result one obtains a maximal value ($=I$) which in fact confirms the result we got at the classical level.

It is instructive to compare the K-amplitudes of the yrast states obtained through diagonalization,
$A^{diag}_{K}$,  and those corresponding to the coherent state (2.3) considered in the minimum point $(\varphi,r)=(0,I)$ for I=20. The latter function can be written in a different form:
\begin{equation}
|\Phi_{IM}\rangle=\left.|\Psi_{IM}\rangle\right|_{0,I}=\sum_{K}\frac{1}{2^I}\left(\begin{matrix}2I\cr I-K\end{matrix}\right)^{1/2}|IMK\rangle \equiv\sum_{K}A^{coh}_K|IMK\rangle.
\label{Phi}
\end{equation}
The two sets of amplitudes were plotted in Fig. 1,
from where we see that the two functions have a similar K dependence. The small difference is caused by the fact that diagonalization provides non-vanishing amplitudes only for $K=even$, while the coherent state  comprises all K-components. The former state is degenerate with the second yrast state which has only $K=odd$ components. Combining the two degenerate functions to a normalized function, the $K$-distribution of the new function is almost identical to that of 
$|\Phi_{IM}\rangle$.
%\end{justify}  
\begin{figure}[ht!]
\begin{minipage}{7cm}
\includegraphics[width=0.7\textwidth]{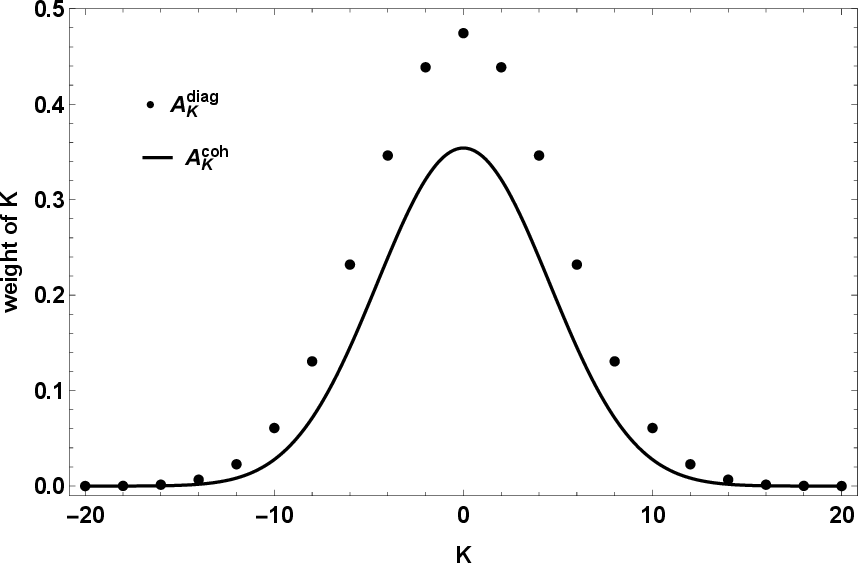}
\end{minipage}\ \ \hspace*{-0.5cm}
\begin{minipage}{7cm}
\caption{The $K$-amplitudes given by the diagonalization procedure and the coherent state, respectively.}
\label{Fig.1}
\end{minipage}
\end{figure}
%\begin{justify}
%%%%%%%%%%%%%%%%%%%%%%%%%%%%%%%%%%%%%%%%%%%%%%%%%%%%%%55
%\end{document}

Having the wave functions for the states I, one can calculate the E2 and M1 transition probabilities using the electric quadrupole and magnetic dipole transition operators: 
\begin{eqnarray}
{\cal M}(E2;\mu)&=&\frac{3}{4\pi}Ze_{eff}R_0^2\left(D^2_{\mu 0}\beta\cos\gamma+(D^2_{\mu 2}+D^2_{\mu , -2})\beta\sin\gamma /\sqrt{2}\right),\nonumber\\
{\cal M}(M1;\mu)&=&\sqrt{\frac{3}{4\pi}}g_RD^1_{\mu\nu} R_{\nu}.
\label{troper}
\end{eqnarray}
Here Z and $R_0$ denote the nuclear charge and radius respectively, while $D^I_{MK}$ stands for the Wigner function describing the rotation matrix, $\mu_N$ is the nuclear magneton and 
$e_{eff}$-
the effective charge. $\beta$ is the nuclear deformation and $\gamma$ represents  the nuclear shape  deviation from the axial symmetry. For the case of $^{158}$Er the nuclear quadrupole deformation is taken equal to 0.203 \cite{Lala} while $\gamma$ is determined using for the moments of inertia the expressions given by the hydrodynamic model on one hand and the values provided by the adopted fitting procedure  for energies, on the other hand. Thus, one obtains:
\begin{equation}
\cot\gamma=\left(-1+2({\cal J}^{hyd}_1/{\cal J}^{hyd}_3)^{1/2}\right)/\sqrt{3}.
\end{equation}
where ${\cal J}^{hyd}_k$ are the moments of inertia corresponding to the k-th axis, respectively, in the liquid drop model \cite{Ring}.
Using the results for the moments of inertia yielded by the fitting procedure one obtains: 
\begin{equation}
\cos\gamma =0.5772,\;\;\sin\gamma=0.8166
\label{sincos}
\end{equation}
 The value extracted from the equations (\ref{sincos}) is about 55$^{0}$ which indicates a shape close to the prolate-oblate transition, i.e. a gamma soft picture.This is reflected in the low part of the spectrum by a even-odd staggering which for large rotation frequency is washed out. Note that the moments of inertia considered in this section are rigid and consequently the corresponding $\gamma$
is not depending on spin. This is at variance with the microscopic description which studied the shape dependence on angular momentum \cite{Lee,Riley,Beck}.For example Ref.\cite{Beck} pointed out that in $^{158}$Er, approaching the band termination the states collectivity decreases which might be caused by the fact that the corresponding shape is almost oblate. Also, the coupling with two or four quasiparticle states may lead to discontinuities in the yrast spectrum 
\cite{Lee,RadBu}. Combination of the single particle level crossing effect and the prolate-oblate competition was accounted for in Ref. \cite{Riley} and a prolate-oblate shape coexistence was pointed out for high spins. Similar result was obtained in Ref.\cite{Shi} when the rotation axis coincides with one principal axis. However, when the rotational axis changes the direction the higher energy minimum becomes a saddle point. Since the approach adopted for $^{158}$Er  yields a energy level sequence which agrees well with the corresponding experimental data as well as with the exact results obtained through diagonalization, it seems that the rigid soft $\gamma$ regime is able to describe the global properties of the wobbling band in $^{158}$Er.

When the intra-band transition is concerned, the initial and final states of the yrast band are described by the function defined in Eq. (\ref{Phi}). Since the one  phonon operator of the yrast band is associated with the quantas in the parameter space, it commutes with the transition operator and moreover gives zero when acts on the final yrast state, unless this deviates from 
(\ref{Phi}) due to parameter fluctuations.
The normalized first order expansion of $\Psi$ around $\Phi$ is:
\begin{eqnarray}
|\Psi_{IM}\rangle &=&N_I\frac{1}{2^I}\sum_{K=-I}^{K=I}\left[1+\frac{i}{\sqrt{2}}\left(\frac{\alpha_I K}{I}
+\frac{I-K}{\alpha_I}\right)\right]\left(\begin{matrix} 2I \cr I-K\end{matrix}\right)^{1/2}|IMK\rangle a_{I}^{\dagger}|0\rangle_I,
\;I\ne.0,\nonumber\\
\left(N_I\right)^{-2}&=&\frac{1}{2^{2I}}\sum_{K=-I}^{K=I}\left[1+\frac{1}{2}\left(\frac{\alpha_I K}{I}+\frac{I-K}{\alpha_I}\right)^2\right] \left(\begin{matrix} 2I \cr I-K\end{matrix}\right).
\end{eqnarray}
Here $a_{I}^{\dagger}$ denotes the creation operator for a wobbling quanta on the top of the yrast state of angular momentum I. The corresponding vacuum state is $|0\rangle_{I}$. The canonical transformation relating the conjugate coordinate and momentum with the creation and annihilation operators depend on the parameter $\alpha_I$ having the expression:
\begin{equation}
\alpha_I=\left(I^2\frac{\frac{1}{{\cal J}_2}-\frac{1}{{\cal J}_1}}{\frac{1}{{\cal J}_3}-\frac{1}{{\cal J}_1}}\right)^{1/4}.
\end{equation}
For what follows we introduce the following notation for a wobbling multi-phonon state:
\begin{equation}
|\Phi_{IM};n_w\rangle = |\Phi_{IM}\rangle \frac{\left(a_{I}^{\dagger}\right)^{n_w}}{\sqrt{n_{w}!}}|0\rangle_I.
\end{equation}

The reduced transition probability is readily obtained:
\begin{equation}
B(E2;In_w\to I'n^{'}_w)=\left|\langle \Phi_I;n_w||{\cal M}(E2)||\Phi_{I'};n^{'}_{w}\rangle\right|^2.
\end{equation}
The transition amplitude can be also used to calculate the quadrupole moment of an yrast state of angular momentum I:
\begin{equation}
Q_{I}=\sqrt{\frac{16\pi}{5}}C^{I\;2\;I}_{I\;0\;I}\langle \Phi_{II}|{\cal M}(E2)|\Phi_{II}\rangle.
\end{equation}
The magnetic properties were studied with the dipole transition operator defined by 
Eq.(\ref{troper}). The result for the magnetic dipole moment for an yrast state $I$ is:
\begin{equation}
\mu_I\equiv\sqrt{\frac{4\pi}{3}}\langle\Phi_{II}|I_0|\Phi_{II}\rangle =\sqrt{\frac{4\pi}{3}}g_RC^{I\;1\;I}_{I\;0\;I}\sqrt{R(R+1)}\mu_N.
\end{equation}
where the standard notation, $\mu_N$, for the nuclear magneton has been used.

These expressions for the electric and magnetic transition operator matrix elements will be used in the next subsection to calculate the corresponding observables for the case of $^{158}$Er.
%%%%%%%%%%%%%%%%%%%%%%%%%%%%%%%%%%%%%%%%%%%%%%%%%%%%%%%%%%%%%%%%%%%%%%%%%%%%%%%%%%%%5

\subsection{Numerical analysis}
Here we address the issue of how do the results obtained through diagonalization, by solving the Schr\"{o}dinger equation and by the harmonic approximation leading to the wobbling motion of the angular momentum respectively, compare with each other. The moments of inertia used here are  those obtained in Ref.\cite{RadBu} by fitting the experimental excitation energies with the wobbling energy formula, i.e. ${\cal I}_1=125\hbar^2MEV^{-1},\; {\cal I}_2=31.4\hbar^2MEV^{-1},\; {\cal I}_3=42 \hbar^2MEV^{-1}$. The variable x from the Bargmann representation of the rotor Hamiltonian is defined in the interval $(-\infty,+\infty)$, while the current variable $t$ entering the Schr\"{o}dinger equation is restricted in a finite interval which is close to $[-1.5,+1.5]$.

%\newpage
\begin{figure}[h!]
\begin{minipage}{6cm}
\includegraphics[height=3cm,width=0.6\textwidth]{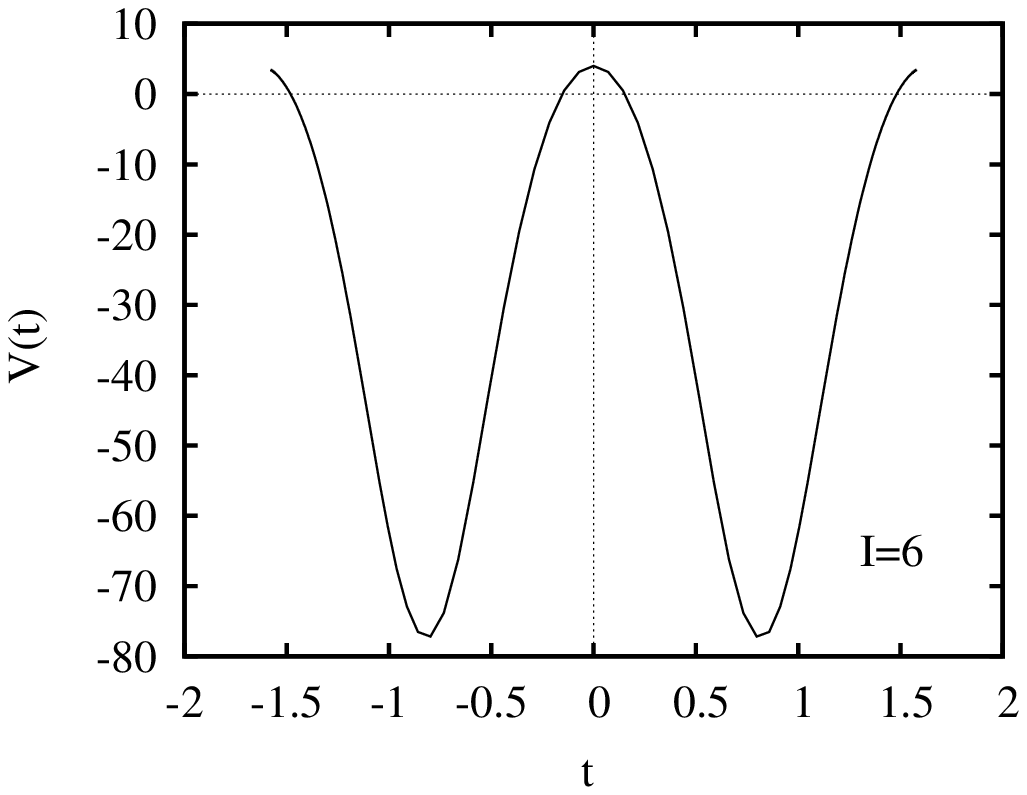}\includegraphics[height=3cm,width=0.4\textwidth]{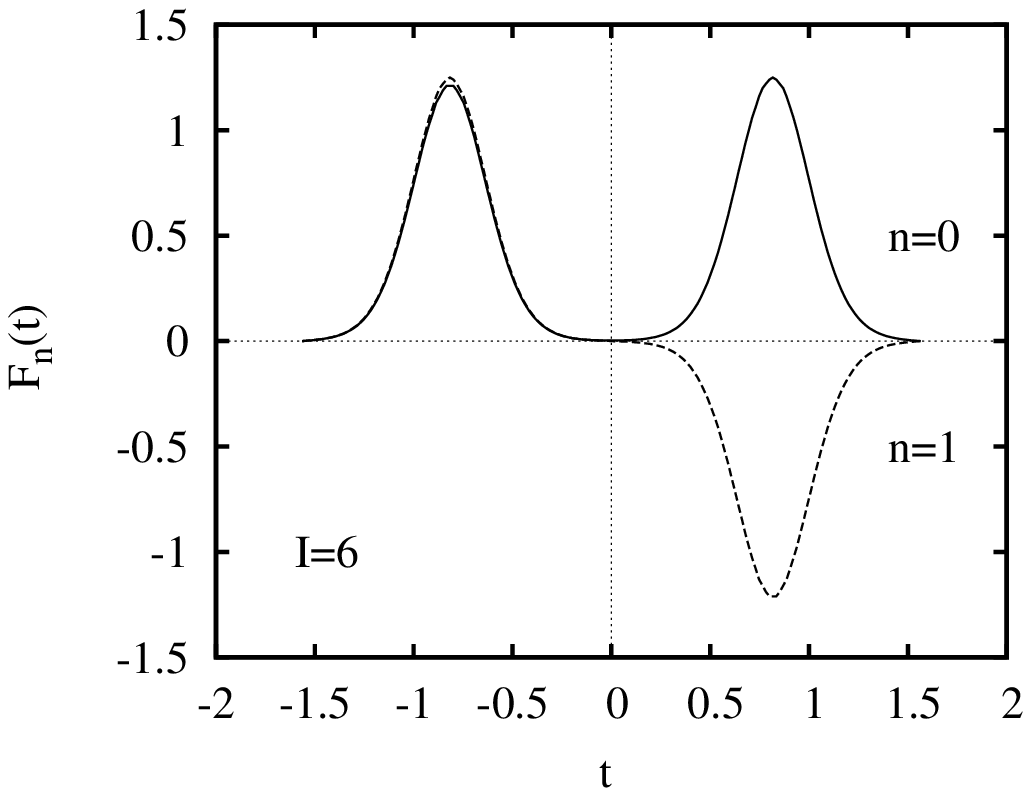}
\end{minipage}\ \
\begin{minipage}{6cm}
\caption{Left panel:The dependence of the potential on the variable t, defined by Eq. (2.38). }
\label{Fig.4}
\caption{Right panel:Eigenfunction $F_n$ as function of t, for n=0,1.}
\end{minipage} 
\label{Fig.5}
\end{figure}
%\end{document}
 In Ref.\cite{Rad07}, the potential energy was considered as function of $x$ , while here its dependence on the variable $t$ is represented in Fig. 2, for I=6. If one calculates the average of $R_1$ with trial function $|\psi(z)\rangle$  and the result is considered in the two potential minima one obtains that $\langle \hat{R}_1\rangle =\pm I.$ This shows that in one minimum the system rotates around the axis OX, while in the other minimum the rotation is performed around -OX. An useful insight to the system behavior, for a given solution of the Schr\"{o}dinger equation, is obtained by plotting the wavefunction $F_n(t)$ for n=0,1;3,4 and 5,6 in Figs.3, 4, 5  respectively, for I=6. The pair of states represented in each of the mentioned figures are degenerate.

\begin{figure}[h!]
\begin{minipage}{6cm}
%\hspace*{-2.5cm}
\includegraphics[height=4cm,width=0.5\textwidth]{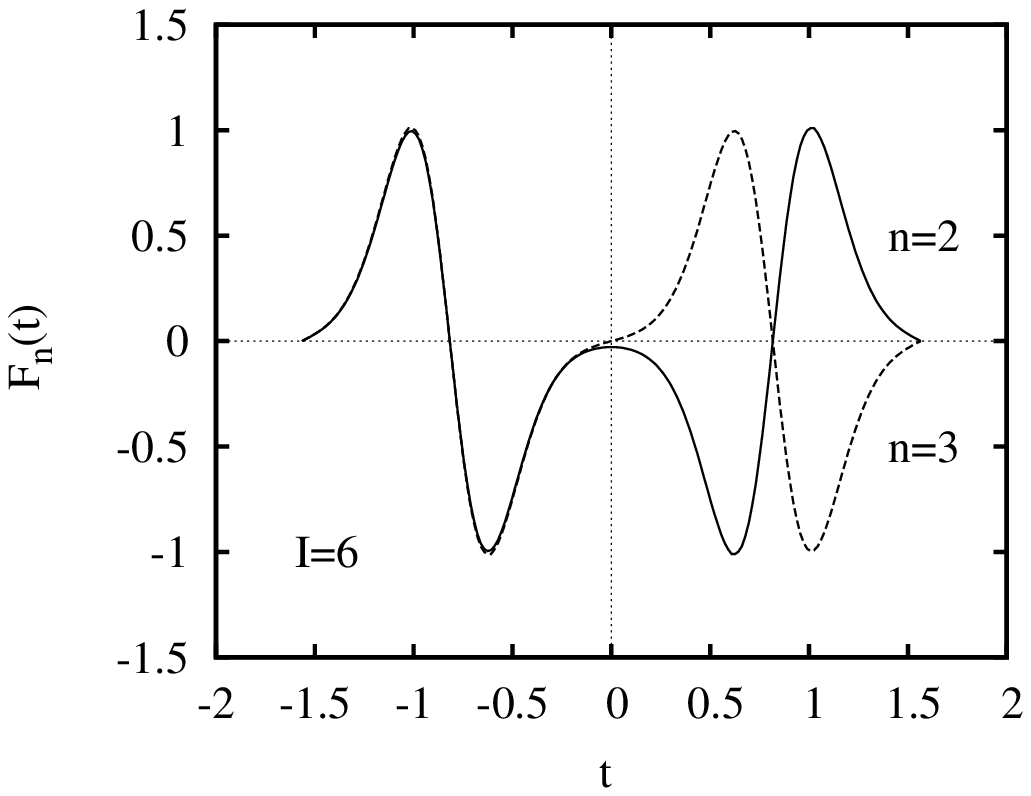}\includegraphics[height=4cm,width=0.5\textwidth]{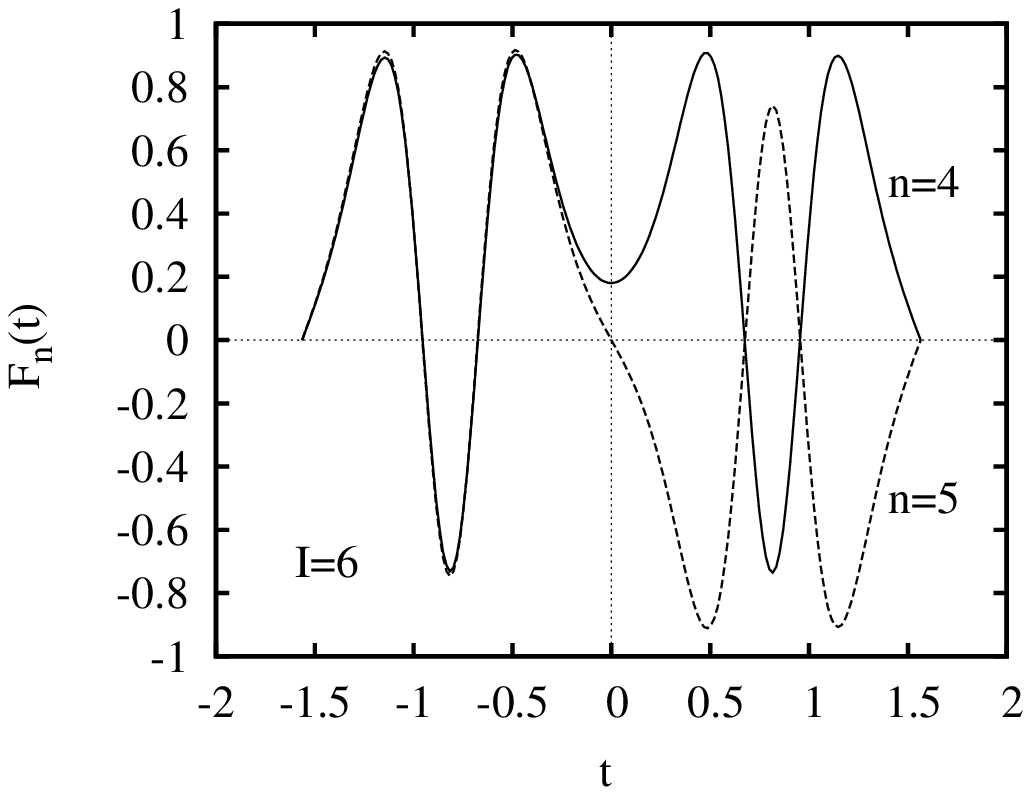}
\end{minipage}\ \
\begin{minipage}{4cm}
%\hspace*{2cm}
\caption{Left panel:The $F_n$ as function of t for n=2,3. }
\label{Fig. 6}
\caption{Right panel:The $F_n$ as function of t for n=4,5. }
\label{Fig.7}
\end{minipage}
\end{figure}

The probability distributions $|F_n|^2$ for the degenerate states  are identical. Note that if the states corresponding to $F_n$ and $F_{n+1}$ are degenerate, then the states described by 
$F_n+F_{n+1}$ and $F_n-F_{n+1}$ are also degenerate and localized each in a separate well.  For an I running from zero to $I_{max}$,  the lowest two eigenstates of the Schr\"{o}dinger equation for each I form two degenerate bands, one localized inside the well corresponding to the positive minimum and one in the well associated to the negative minimum. The same is also true for the next two degenerate bands and so on.

\begin{figure}[h!]
\begin{minipage}{6cm}
\includegraphics[height=4cm,width=0.5\textwidth]{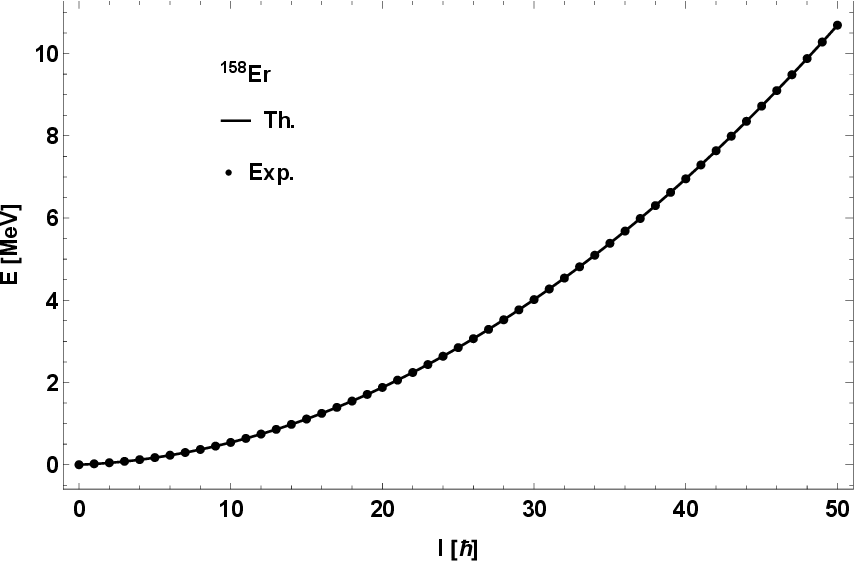}\includegraphics[height=4cm,width=0.5\textwidth]{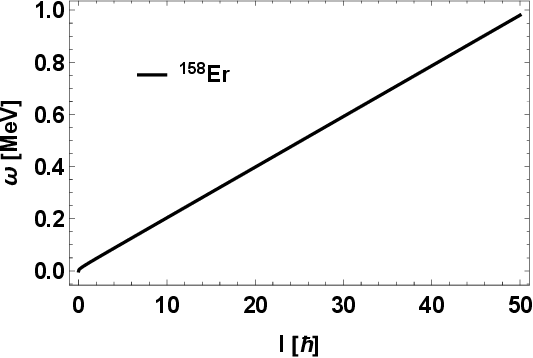}
\end{minipage}\ \
\begin{minipage}{6cm}
\caption{Left panel: Results (Th.) obtained with Eq.(2.29) are compared with experimental data (Ex.) taken from \cite{Nica}.}
\label{yrer}
\caption{Right panel:The wobbling frequency for $^{158}$Er as function of the angular momentum.}
\label{wober}
\end{minipage}
\end{figure} 
The parameters ${\cal I}_k$ used in our numerical analysis are those obtained by fitting the experimental yrast energies for $^{158}$Er with the relation (\ref{yra}).

We calculated the yrast energies provided by three methods: diagonalization, solving the Schr\"{o}dinger equation and by the wobbling energy formula (2.29). The results is that the three sets of energies are almost equal to each other, the maximal deviations being less that 5 keV.

It is known that for triaxial nuclei, i.e. when the three moments of inertia are all different, the projection of ${\bf R}$ on the OZ axis is not a good quantum number. The present result shows that the coherent states we used as trial function is an optimal mixture of the K components to approximate the exact wavefunction. Also, the wobbling approximation of the yrast energies describes very well the exact solution of the Schr\"{o}dinger equation.
%\end{document}
The agreement of th calculated yrast wobbling energies and  the corresponding experimental data is very good as shown in Fig.\ref{yrer}. Theoretical results were obtained with the wobbling formula (2.29). We notice that the wobbling frequency depends almost linearly on the angular momentum I. This dependence can be seen in Fig.\ref{wober}. 

\begin{table}
\begin{tabular}{|c|cc|c|c|c|c|c|}
\hline
&\multicolumn{3}{c|}{$B(E2;In_w\to (I-2+n_w)n^{'}_w)$} &$B(E2;In_w\to I-1n^{'}_w)$&
$B(M1;In_w\to I-1n^{'}_w)$&$Q_I$ & $\mu_{I}$\\
&\multicolumn{3}{c|}{$[W.u]$}    &  $[W.u]$&$[10^{-4}\mu_N^2]$&$[e.fm^2]$&$[\mu_N]$\\
\hline
&\multicolumn{2}{c|}{$n_w=n^{'}_w=0$} & $n_w=1\;n^{'}_w=0$ & $n_w=1\;n^{'}_w=0$&$n_w=1\;n^{'}_w=0 $      &             &   \\
I& Th.         &       Exp. \cite{Nica}    &    Th.                & Th.       &  Th. &Th.   &Th.          \\
\hline
2&   61.287    &      129$\pm $9               &     0    &0.916&0.05           & 73.627    &  0.861  \\       
4&  170.243    &      186$\pm$ 6                &  75.961 &0.453&0.250&131.190        &     1.721           \\
6&  212.149    &      246$\pm$ 8                &  130.908&0.258&0.420&161.979   &   2.582        \\
8&  234.334    &      298$\pm$ 10                &  165.756&0.165&0.511&180.838        &3.443  \\
10& 248.068    &      250$\pm$ 4                &  189.322 &0.114&0.558& 193.525     & 4.304                 \\
12& 257.407    &      260$\pm$3                &  206.216 &0.084&0.585& 202.631     &5.164                  \\
\hline
\end{tabular}
\caption{\textmd{The calculated intra-band B(E2) values are compared with the available experimental data, taken from Ref.\cite{Nica}.Also, the calculated B(M1) values connecting the $\Delta I=1$ yrast states as well the theoretical values for the quadrupole and magnetic moments are listed.}}
\label{Table 1.}
\end{table}

Results for transition probabilities and moments are obtained with an effective charge $e_{eff}=1.4 e$ and collected in Table I. The agreement with the corresponding experimental data for the B(E2) values is good.
%\end{document}
Concluding, the semi-classical formalism presented in this section describes the experimental data in a realistic fashion. We showed that the variational equations corresponding to a coherent
 state as trial function provides, after re-quantization, a very good description of the quantal system.

\section{Description of the wobbling motion in even-odd nuclei}  
\renewcommand{\theequation}{8.3.\arabic{equation}}
\setcounter{equation}{0}
We suppose that the odd-mass nuclear system consists of an even-even core described by a triaxial rotor Hamiltonian and a single j-shell particle moving in a quadrupole deformed mean-field described by:
\begin{equation}
H_{sp}=\epsilon_j+\frac{V}{j(j+1)}\left[\cos\gamma(3j_3^2-{\bf j}^2)-\sqrt{3}\sin\gamma(j_1^2-j_2^2)\right].
\end{equation}
where $\epsilon_j$ denotes the single particle energy of the odd nucleon.
It is convenient to express the rotor Hamiltonian in terms of the total angular momentum ${\bf I}$ and the angular momentum carried by the odd particle:
\begin{equation}
H_{rot}=\sum_{k=1,2,3}A_k(I_k-j_k)^2.
\end{equation}
Where $A_k$ are expressed in terms of the moments of inertia associated to the principal axes of the inertia ellipsoid as:$A_k=\frac{1}{2{\cal I}_k}$.

In what follows, the moments of inertia are taken as given by rigid-body model in the Lund convention:
\begin{equation}
{\cal I}^{rig}_{k}=\frac{{\cal I}_0}{1+(\frac{5}{16\pi})^{1/2}\beta}\left[1-\left(\frac{5}{4\pi}\right)^{1/2}\beta\cos\left(\gamma+\frac{2}{3}\pi k\right)\right],\;k=1,2,3
\end{equation}

To the total Hamiltonian, $H (=H_{rot}+H_{sp})$,
we associate the time dependent variational equation
\begin{equation}
\delta\int_{0}^{t}\langle \Psi|H-i\frac{\partial}{\partial t'}|\Psi\rangle d t'=0,
\end{equation}
where the trial function is chosen as:
\begin{equation}
|\Psi\rangle ={\bf N}e^{z\hat{I}_-}e^{s\hat{j}_-}|IMK\rangle |jj\rangle,
\end{equation} 
with $\hat{I}_-$ and $\hat{j}_-$ denoting the lowering operators for the intrinsic angular momenta ${\bf I}$ and ${\bf j}$ respectively, while ${\bf N}$ is  the normalization factor having the expression:
\begin{equation}
{\bf N}^{-2}=(1+|z|^2)^{2I}(1+|s|^2)^{2j}.
\end{equation}
The variable $z$ and $s$ are complex functions of time and play the role of classical phase space coordinates describing the motion of the core and the odd particle, respectively:
\begin{equation}
z=\rho e^{i\varphi},\;\;s=fe^{i\psi},
\end{equation}
The variables $(\varphi, r)$ and $(\psi,t)$ with $r$ and $t$ defined as:
\begin{equation}
r=\frac{2I}{1+\rho^2},\;\;0\le r\le 2I,\;\;\;t=\frac{2j}{1+f^2},\;\; 0\le t\le 2j,
\end{equation}
bring the classical equations provided by the variational principle to the canonical form:
\begin{equation}
\frac{\partial {\cal H}}{\partial r}=\stackrel{\bullet}{\varphi};\;\;\frac{\partial {\cal H}}{\partial \varphi}=-\stackrel{\bullet}{r};\;\; \frac{\partial {\cal H}}{\partial t}=\stackrel{\bullet}{\psi};\;\;\frac{\partial {\cal H}}{\partial \psi}=-\stackrel{\bullet}{t}. 
\label{eqmot}
\end{equation}
where ${\cal H}$ denotes the average of $H$ (Eq.3.5) with the function $|\Psi\rangle$ and has the expression:
\begin{eqnarray}
{\cal H}&=&\frac{I}{2}(A_1+A_2)+A_3I^2+\frac{2I-1}{2I}r(2I-r)\left(A_1\cos^2\varphi+A_2\sin^2\varphi -A_3\right)\nonumber\\
        &+&\frac{j}{2}(A_1+A_2)+A_3j^2+\frac{2j-1}{2j}t(2j-t)\left(A_1\cos^2\psi+A_2\sin^2\psi -A_3\right)\nonumber\\
        &-&\sqrt{r(2I-r)t(2j-t)}\left(A_1\cos\varphi\cos\psi+A_2\sin\varphi\sin\psi\right)+A_3\left(r(2j-t)+t(2I-r)\right)\nonumber\\
        &+&V\frac{2j-1}{j+1}\left[\cos\gamma-\frac{t(2j-t)}{2j^2}\sqrt{3}\left(\sqrt{3}\cos\gamma -\sin\gamma\cos2\psi\right)\right].
\label{energyfunction}
\end{eqnarray}
From Eq.(\ref{eqmot}) we see that the angles $\varphi$ and $\psi$ play the role of generalized coordinates while $r$ and $t$  are the corresponding conjugate momenta.
Looking for the extremal points of the energy surface ${\cal H}=const$, one finds out that the point $(\varphi,r;\psi,t)=(0,I;0,j)$ is a minimum point for the classical energy function.
Linearizing the equations of motion around the mentioned minimum point, one arrive at a homogeneous and linear system for the generalized conjugate coordinates. The compatibility restriction leads to an equation defining the wobbling frequency:
\begin{equation}
\Omega^4+B\Omega^2+C=0,
\label{equOm}
\end{equation}
where the coefficients B and C have the expressions:
\begin{eqnarray}
-B&=&\left[(2I-1)(A_3-A_1)+2jA_1\right]\left[(2I-1)(A_2-A_1)+2jA_1\right]+8A_2A_3Ij\nonumber\\
 &+&\left[(2j-1)(A_3-A_1)+2IA_1+V\frac{2j-1}{j(j+1)}\sqrt{3}(\sqrt{3}\cos\gamma+\sin\gamma)\right]\nonumber\\
 &\times&\left[(2j-1)(A_2-A_1)+2IA_1+V\frac{2j-1}{j(j+1)}2\sqrt{3}\sin\gamma\right],\\
C&=&\left\{\left[(2I-1)(A_3-A_1)+2jA_1\right]\left[(2j-1)(A_3-A_1)+2IA_1+V\frac{2j-1}{j(j+1)}\sqrt{3}(\sqrt{3}\cos\gamma+\sin\gamma)\right]-4IjA_3^2\right \}\nonumber\\
 &\times&\left\{\left[(2I-1)(A_2-A_1)+2jA_1\right]\left[(2j-1)(A_2-A_1)+2IA_1+V\frac{2j-1}{j(j+1)}2\sqrt{3}\sin\gamma\right]-4IjA_2^2\right\}.\nonumber\\
\label{BandC}
\end{eqnarray}

Under certain restrictions for MoI's, the dispersion equation (\ref{equOm}) admits two real and positive solutions. Hereafter, these will be denoted by $\Omega^{I}_1$ and $\Omega^{I}_{2}$ for $j=i_{13/2}$, with $\Omega^{I}_1 < \Omega^{I}_{2}$.

 In Ref.\cite{Rad017,RaPoAl}the wobbling frequencies have been used in order to define the rotational bands in the even-odd isotopes $^{161,163,165,167}$Lu. Most experimental available data are in $^{163}$Lu, where four bands denoted by TSD1, TSD2, TSD3 and  TSD4 are known. The name is the achronime for "triaxial superdeformed", which reflects the fact that the nuclear deformation and the deviation from the axial symmetric shape  are large ($\beta\approx 38$ and $\gamma\approx 20^{0}$).  In the quoted paper the variational principle is used only for the states from TSD1 while for the three bands are interpreted as one, two  and three boson excitations of the ground band. The ground band excitation energies are just the zero point energies associated to each member of the band. 
The band TSD4 is known only for $^{163}$Lu and has a negative parity in contrast with the other bands which are of positive parity. The single particle orbital for the odd nucleon was taken as $i_{13/2}$ in the first three bands, while for TSD4 that is $h_{9/2}$. Due to the core polarization caused by the particle-core interactions the the core moments of inertia (MoI) for 
TSD1,TSD2 and TSD3 are the same but different from those characterizing TSD4. A good description of the data was obtained although due to situation mentioned in the above statement the fitting procedure is somewhat tedious.

Here we give the results of Ref.\cite{PoRa,Rad20} where variational principle is applied to the states from TSD1, TSD2 and TSD4, while TSD3 is considered to be an one phonon excitation of TSD2. Moreover for all the four bands the odd nucleon is moving in the orbital $i_{13/2}$. The core states are given by the triaxial rotor and are of positive parity for TSD1, TSD2 and TSD3 and of negative parity for TSD4. Note the fact that the pair of bands TSD1, TSD3 and TSD2, TSD4 have the same signature, $+1/2$ and $-1/2$, respectively. In this context the bands TSD2 and TSD4 may be called {\it parity partner bands}.

Further, to the $TSD_{1,2,3,4}$ bands we associate the energies:

\begin{eqnarray}
  &&  E_I^\text{TSD1}=\epsilon_{j} + \mathcal{H}_\text{min}^{(I,j)}+\mathcal{F}_{00}^I , \;\;I=R+j, R=0,2,4,. . . ,\nonumber \\
  &&  E_I^\text{TSD2}=\epsilon_{j,1} + \mathcal{H}_\text{min}^{(I,j)}+\mathcal{F}_{00}^I , \;\;I=R+j, R=1,3,5,. . . \nonumber\\
  &&  E_I^\text{TSD3}=\epsilon_{j} + \mathcal{H}_\text{min}^{(I,j)}+\mathcal{F}_{10}^I, \;\; I=R+j, R=0,2,4,. . .  \nonumber\\
  &&  E_I^\text{TSD4}=\epsilon_{j,2} + \mathcal{H}_\text{min}^{(I,j)}+\mathcal{F}_{00}^I,\;\;I=R+j,\;R=1,3,5,. . . . \nonumber\\ \label{wobbling_energies}
\end{eqnarray}

where $\mathcal{F}_{n_{w_1}n_{w_2}}$ is function of the wobbling frequencies

\begin{align}
    F_{n_{w_1}n_{w_1}}^I=(n_{w_1}+\frac{1}{2})\Omega_1^I+(n_{w_2}+\frac{1}{2})\Omega_2^I.  \label{phonons}
\end{align}
while $\mathcal{H}^{(I,j)}_{min}$  is the minimal classical energy.  We considered different re-normalizations for the single-particle mean field in the signature unfavored as well as in the negative parity states, which result two distinct energy shifts for the excitation energies in the TSD2 and TSD4 bands, respectively. These two quantities will be adjusted throughout the numerical calculations such that the energy spectrum is best reproduced. In order to save the space only the results for $^{163}$Lu are presented here. The fitting procedure yields se set $\mathcal{P}=({\cal I}_1,{\cal I}_2,{\cal I}_3)$ shown in the next Table.
\begin{table}[h]
    \centering
  \begin{tabular}{lllll}
  \hline
$\mathcal{I}_1$ [$\hbar^2$/MeV] & $\mathcal{I}_2$ [$\hbar^2$/MeV]& $\mathcal{I}_3$ [$\hbar^2$/MeV] & $\gamma$ [deg. ] & $V$ [MeV] \\
\hline
\hline
72              & 15              & 7               & 22       & 2. 1
\end{tabular}
    \caption{\textmd{The parameter set $\mathcal{P}$ that was determined by a fitting procedure of the excitation energies of $^{163}$Lu. }}
    \label{parameter_set}
\end{table}

\begin{figure}
    \centering
    \includegraphics[scale=0.30]{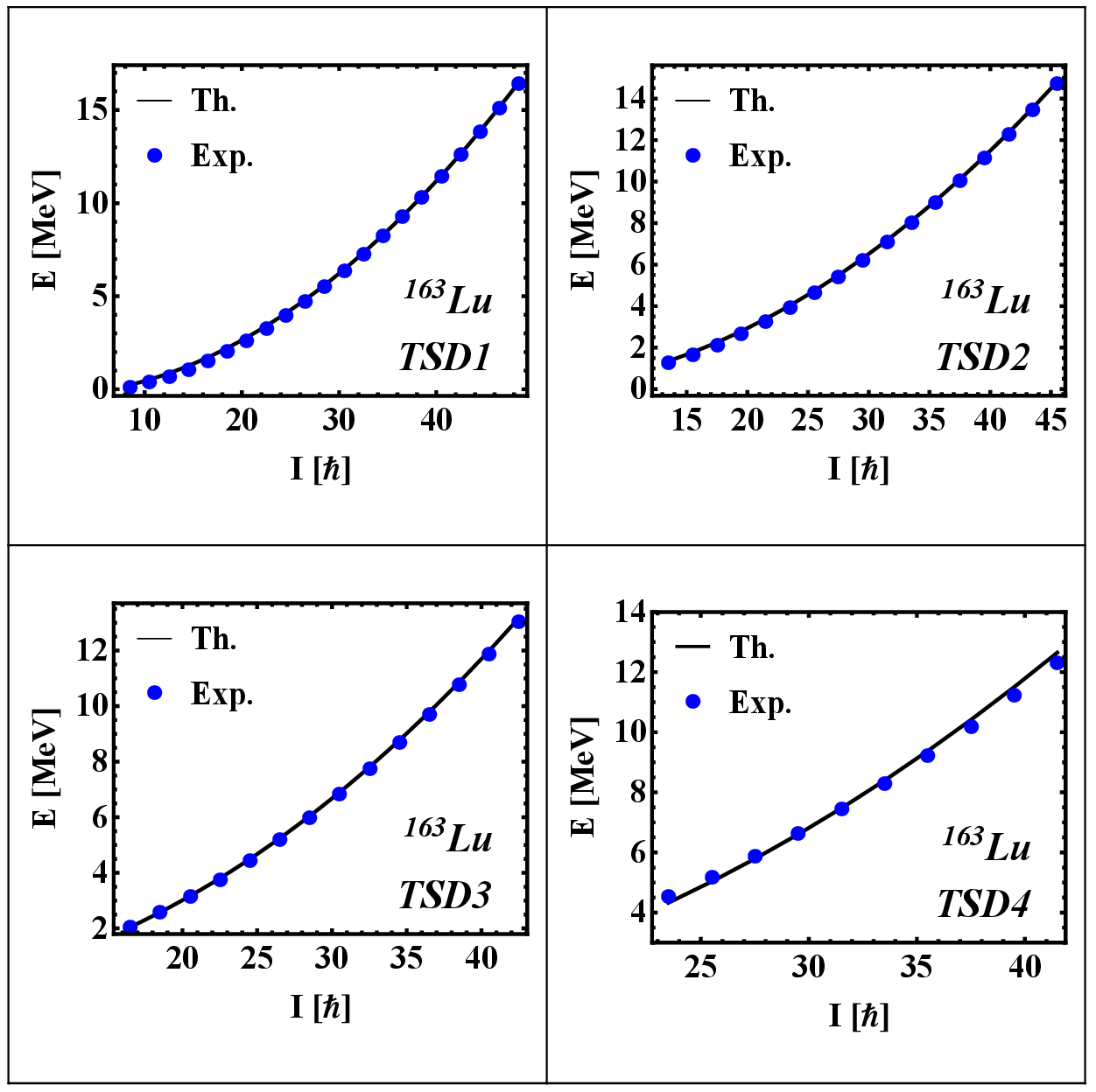}
    \caption{The excitation energies for the bands TSD1, TSD2, TSD3, and TSD4. }
    \label{tsd_bands}
\end{figure}
Using these parameters the excitation energies in the four bands were calculated and compared with he corresponding experimental data, as shown in Fig. 9.
In terms of the stability of the wobbling motion with respect to the total angular momentum, several contour plots were plotted, using the obtained parameter set $\mathcal{P}$ with the help of Eq.  \ref{energyfunction}.  For each band, a spin close to the band head of each sequence was chosen.  Due to the obtained MOI ordering, the surfaces have minimum points indicated by the red dots for each figure.  Results can be seen in Figs.  \ref{contour-tsd1},\ref{contour-tsd3}.  The four figures have many similarities suggesting common collective properties, but also differences caused by the fact that minima have different depths. The common feature consists of that the equi-energy curves surround a sole minimum for low energy while for higher energies the trajectories go around all minima, the lack of localization indicating an unstable picture. 

\begin{figure}
    \centering
\includegraphics[scale=0.25]{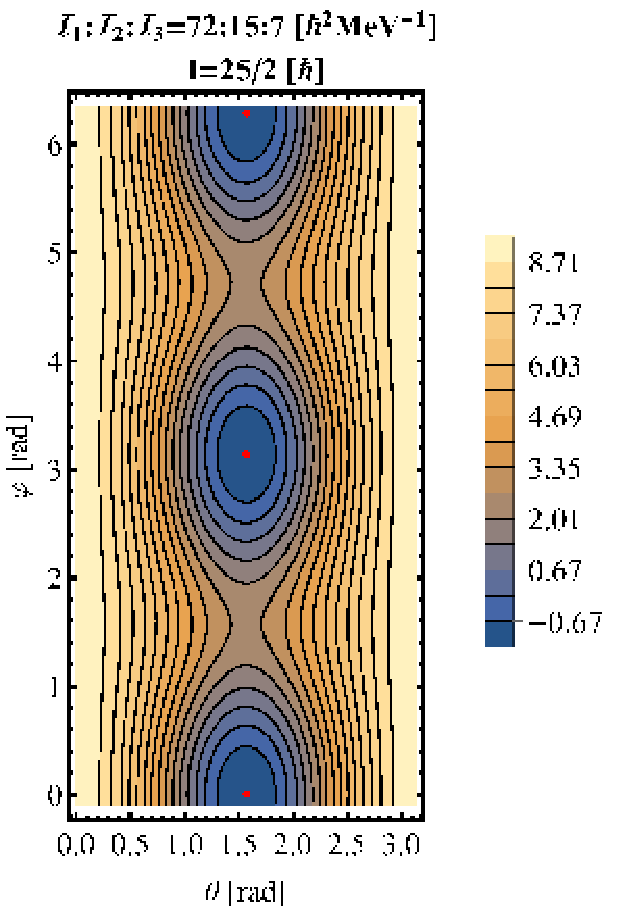}\includegraphics[scale=0.25]{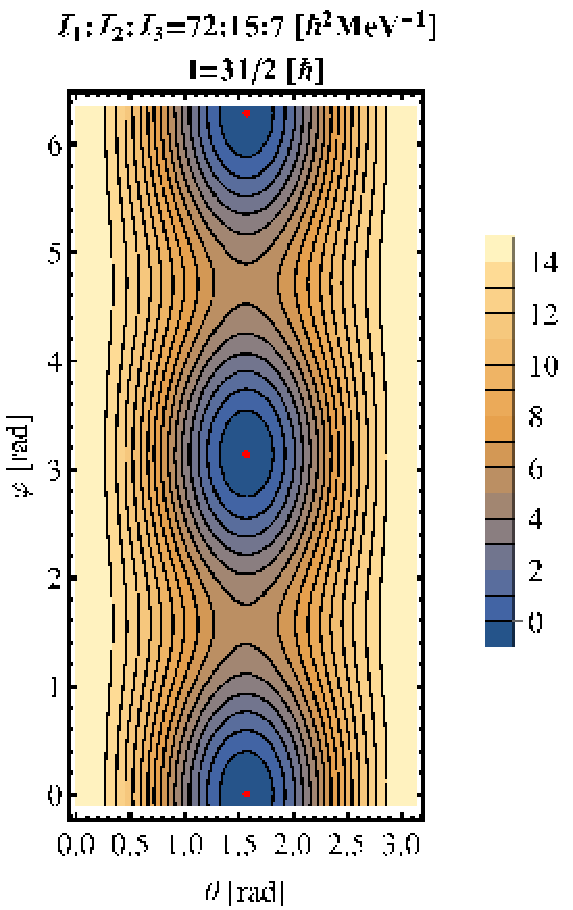}\includegraphics[scale=0.25]{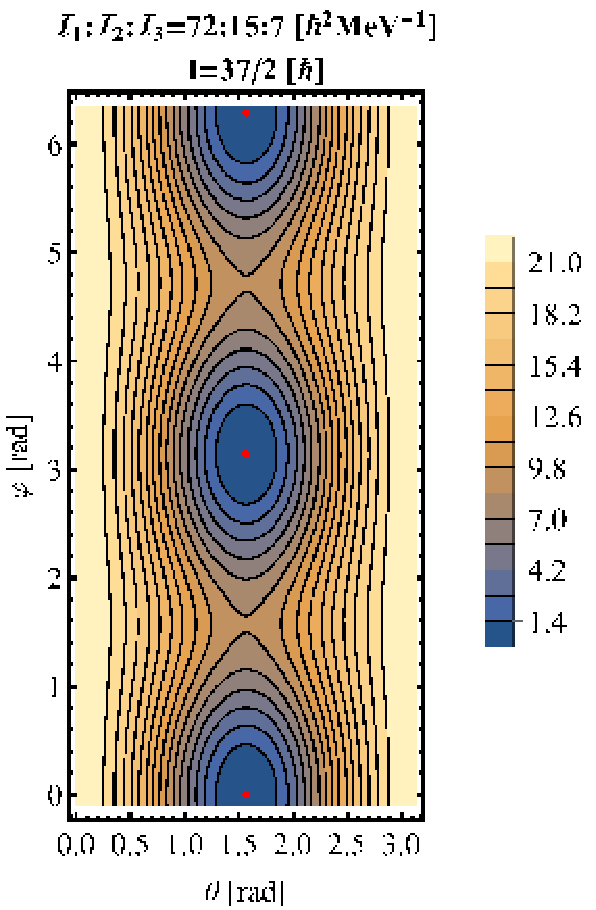}\includegraphics[scale=0.25]{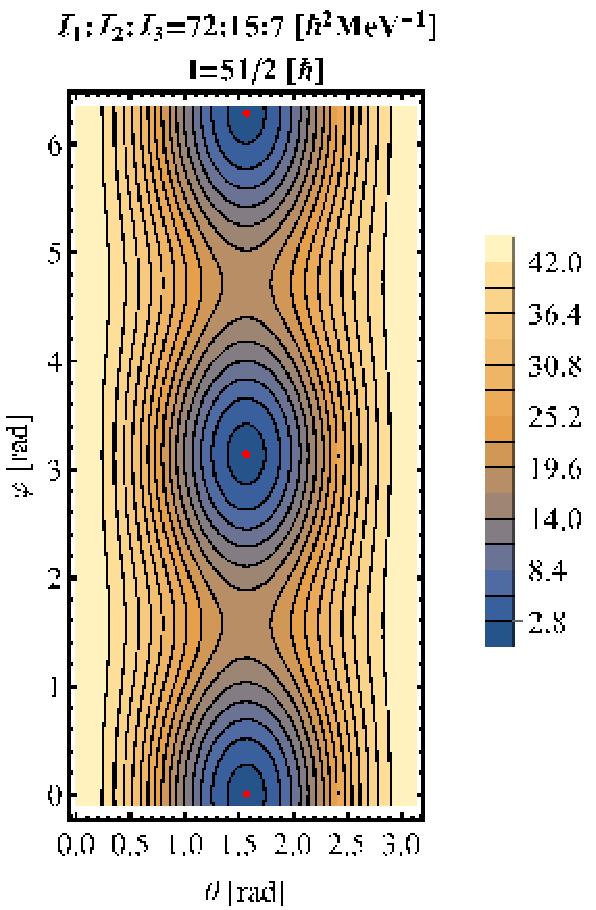}
    \caption{A contour plot with the energy function $\mathcal{H}$ for TSD1, TSD2, TSD3 and TSD4.  The parameter set $\mathcal{P}$ was used for the numerical calculations. }
    \label{contour-tsd1}
\end{figure}

Finally we are interested in finding out the dependence of the classical trajectories on angular momenta as well as on energies.  Indeed, when the model Hamiltonian is diagonalized for a given I,  a set of $2I+1$ energies are obtained.  Therefore, it makes sense to study the trajectory change at increasing the energy. Trajectories are represented in Fig. 11 as the manifold given by intersecting the surfaces corresponding to the two constants of motion, the energy and the angular momentum.  The first energy in each row corresponds to the real excitation energy  for that particular spin state, the second one represents the point at which the ellipsoid touches the sphere at the equator, which marks a nuclear phase transition - while the third one is the trajectory of the system at energies sufficiently large that the system changes its wobbling regime.  For low energies, one notices two distinct trajectories having as rotation axes the 1-axis and -1-axis, respectively.  As energy increases the two trajectories approaches each other which results a tilted rotation axis for each of trajectories, the rotation axes being dis-aligned.  Note that this picture is fully consistent with that of Ref. \cite{Lawr}. When the two trajectories intersect each other, the trajectories surround both minima.  Increasing the energy even more one arrives again at two trajectories regime but with different rotation axes which become close to the 3-axis.  This reflects another phase transition for the system. 

\begin{figure}
%\begin{minipage}{7cm}
%    \centering
    \includegraphics[scale=0.45]{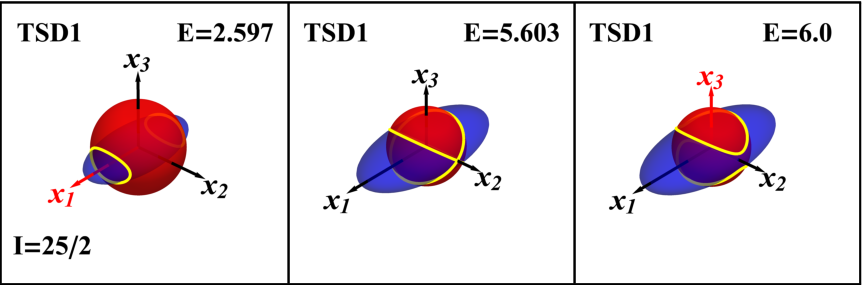}\includegraphics[scale=0.45]{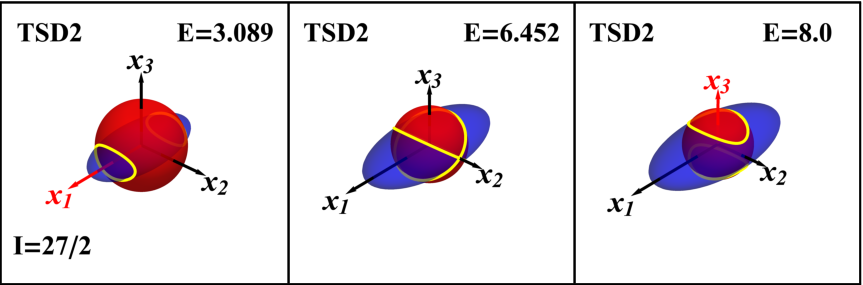}
    \includegraphics[scale=0.45]{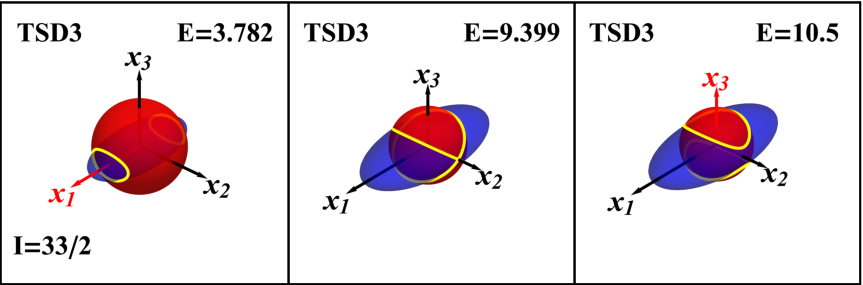}\includegraphics[scale=0.45]{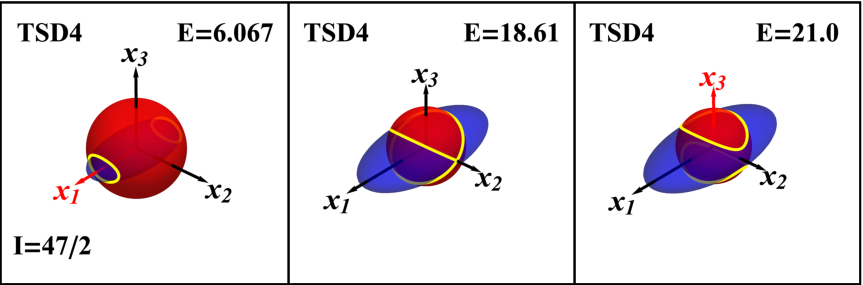}
%\end{minipage}\ \
%\begin{minipage}{7cm}
    \caption{The nuclear trajectory of the system for a spin state belonging to each of the four TSD bands of $^{163}$Lu.  Intersection line marked with yellow color represents the actual orbits. }
%\end{minipage}
    \label{ellipsoids-tsd1}
\end{figure}
\vspace*{-0.2cm}
\subsection{Another scenarios}
An alternative way to interpret the four bands in $^{163}$ is to keep the definition for TSD1, TSD2 and TSD3 as before, but as concerns TSD4 the odd nucleon is moving in the single shell
 j=h$_{9/2}$, otherwise the variational principle is formulated for each state of the negative parity band.
The wobbling frequencies will be denoted by $\Omega^{I}_1$ and $\Omega^{I}_{1'}$ for $j=i_{13/2}$,
and by $\Omega_2$ and $\Omega_{2'}$ for $j=h_{9/2}$. They are ordered as: $\Omega^I_1<\Omega^I_{1'}$ and $\Omega^I_2<\Omega^I_{2'}$. Energies of the states in the four bands are defined as:
\begin{eqnarray}
E^{TSD1}_I&=&\epsilon_{13/2}+{\cal H}_{I,min}(13/2)+\frac{1}{2}\left(\Omega^I_1+\Omega^I_{1'}\right),\nonumber\\
&&I=13/2, 17/2, 21/2,.....\nonumber\\
E^{TSD2}_I&=&\epsilon_{13/2}+{\cal H}_{I,min}(13/2)+\frac{1}{2}\left(\Omega^I_1+\Omega^I_{1'}\right),\nonumber\\
&&I=27/2, 31/2, 35/2,.....\nonumber\\
E^{TSD3}_I&=&\epsilon_{13/2}+{\cal H}_{I-1,min}(13/2)+\frac{1}{2}\left(3\Omega^{I-1}_1+\Omega^{I-1}_{1'}\right),\nonumber\\
&&I=33/2, 37/2, 41/2,.....\nonumber\\
E^{TSD4}_I&=&\epsilon_{9/2}+{\cal H}_{I,min}(9/2)+\frac{1}{2}\left(\Omega^I_2+\Omega^I_{2'}\right),\nonumber\\
&&I=47/2, 51/2, 55/2,.....
\label{ener}
\end{eqnarray}

The excitation energies are obtained by subtracting $E^{TSD1}_{13/2}$ from the above expressions. The  excitation energies for the TSD4 states contain the constant term $\epsilon_{9/2}-\epsilon_{13/2}=-0.334$ MeV. 
 The involved parameters are fitted through the least mean square procedure with the result shown in Table 3.

\begin{table}[h!]
\begin{tabular}{|c|c|c|c|c|c|c|c|c|c|}
\hline
isotope&j& bands& ${\cal I}_1$& ${\cal I}_2$ &${\cal I}_3$& V &$\gamma $ &nr &r.m.s.\\
       & &      & $[\hbar^2/MeV]$&$[\hbar^2/MeV]$&$[\hbar^2/MeV]$& [MeV]&  [deg]   &  states  &[MeV]\\
\hline
$^{163}$Lu&13/2&TSD1,TSD2,TSD3&63.2  &  20  &  10  & 3.1 &  17&52&0.264\\
          &9/2&TSD4           &67    &  34.5 & 50  & 0.7&  17&10&0.057\\
\hline
\end{tabular}
\caption{\textmd{The MoI's, the strength of the single particle potential (V), and  the triaxial parameter ($\gamma$) as provided by the adopted fitting procedure.}}
\label{Table 3}
\end{table}
As seen from Fig.12 the fitted parameters provide excitation energies which agree very well with the corresponding experimental data.

\begin{figure}[h!]
\begin{center}
\includegraphics[height=4cm,width=0.3\textwidth]{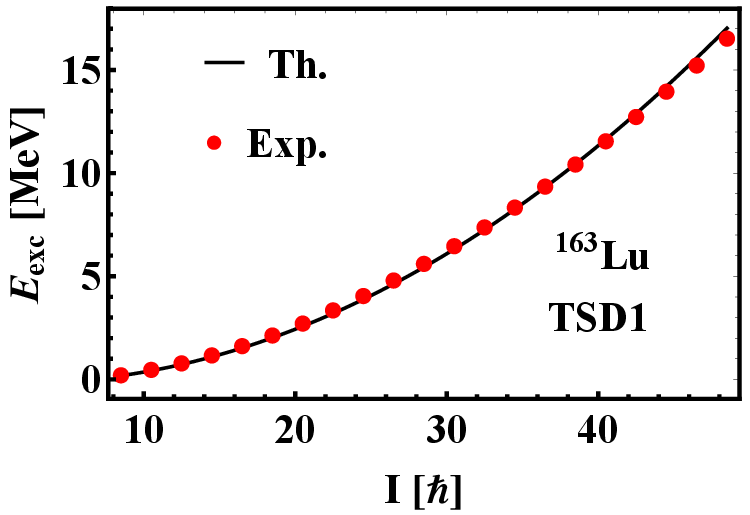}\includegraphics[height=4cm,width=0.3\textwidth]{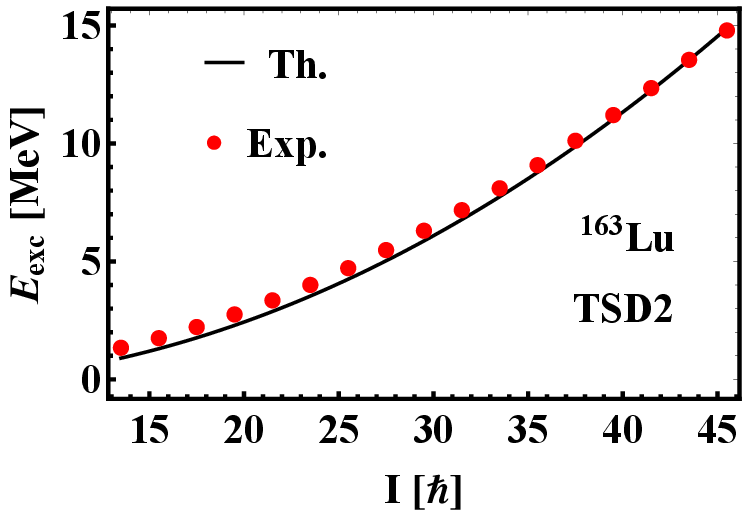}\includegraphics[height=4cm,width=0.3\textwidth]{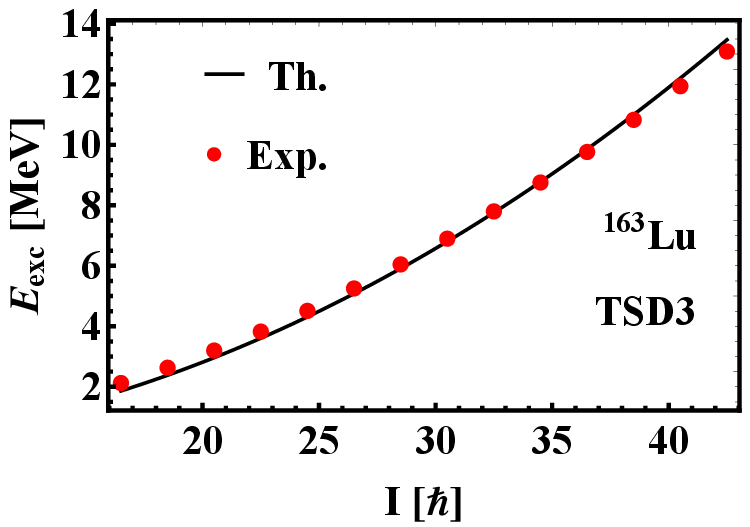}\includegraphics[height=4cm,width=0.3\textwidth]{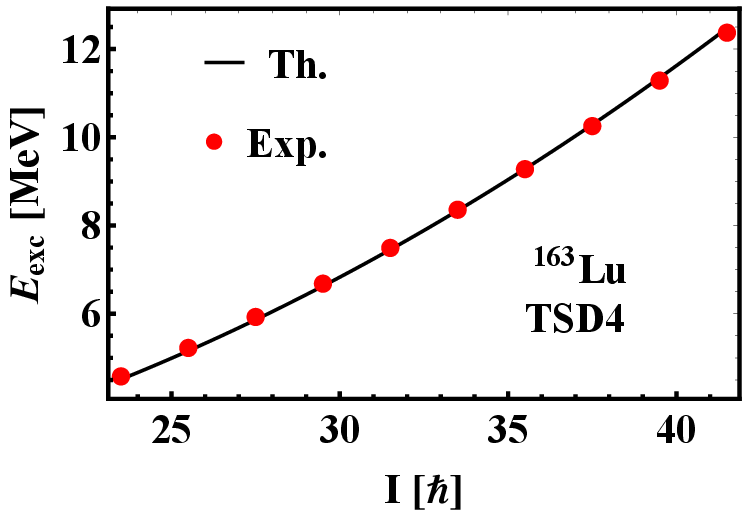}
\caption{Calculated energies for the bands TSD1, TSD2, TSD3 and TSD4 are compared with the corresponding experimental data \cite{Jens1,Hage} for $^{163}$Lu.}
\label{Fig.3}
\end{center}
\end{figure}

The adopted option of fitting the MoI's is naturally required by the observation that the experiment indicates that they are neither irrotational nor rigid but satisfy the relation
${\cal I}_{1}^{irr}<{\cal I}_{1}<{\cal I}_{1}^{rig}$.
Note the MoI's provided by the fitting procedure take care of the particle-core interaction. In that respect the fact that the maximal MoI is ${\cal I}_{2}$ does not necessarily imply that the system motion is of transversal character. Indeed, even though for the non-interacting core the ordering is ${\cal I}_{2}>{\cal I}_{1}>{\cal I}_{3}$, as the microscopic studies show, the particle-core interaction renormalizes the bare MoI's which results in a strong increasing of ${\cal I}_{1}$ (due to the alignment) and only a moderate decreasing of ${\cal I}_{2}$ (caused by the pairing interaction), ending with the dominance of the new ${\cal I}_{1}$, characterizing the whole system \cite{Matsu}. Then, one can assert that the interaction with the odd proton stabilizes the system into a large deformed shape and moreover drives it to a longitudinal-like motion where the maximal MoI is the normalized ${\cal I}_{1}$. This change in the rotation regime is caused by both the angular momentum alignment and the pairing interaction. The transition from a transversal to a longitudinal wobbling motion is not abruptly achieved, but only at a certain critical angular momentum $I_{cr}$. Note that the MoI's were fixed such that the best agreement with the corresponding experimental data is obtained for energies of the whole spectrum and therefore no angular momentum dependence can be inferred. Furthermore, the study of the phase transition, transversal-longitudinal, cannot be performed with the present formalism. However, due to this feature one may say that the present formalism does not exclude the possible transversal wobbling motion in the low lying spectrum where the alignment is small, but that part of bands cannot be explored because the adopted fitting procedure does not use any I dependence for  MoI's.

The fitted parameters yield the excitation energies in the four bands, which are compared with the corresponding experimental energies as shown in Fig. 12.
\begin{figure}[ht!]
\begin{center}
\includegraphics[height=4cm,width=0.3\textwidth]{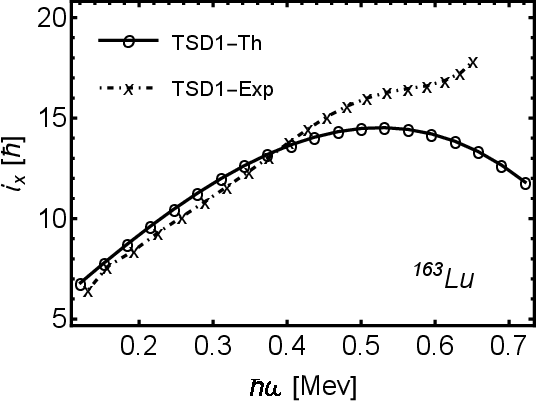}\includegraphics[height=4cm,width=0.3\textwidth]{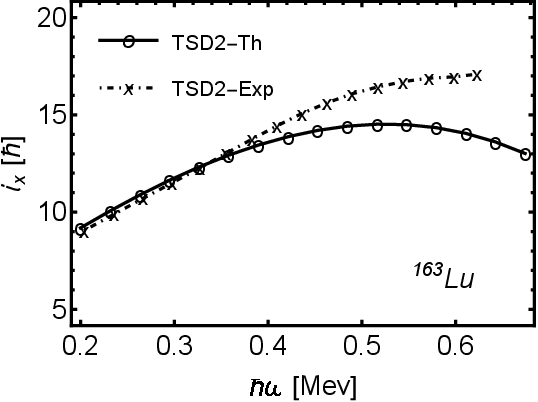}\includegraphics[height=4cm,width=0.3\textwidth]{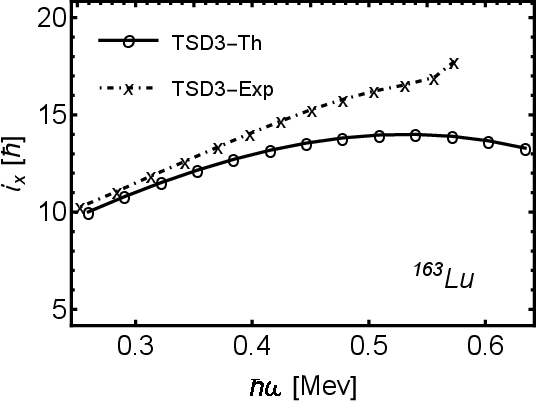}\includegraphics[height=4cm,width=0.3\textwidth]{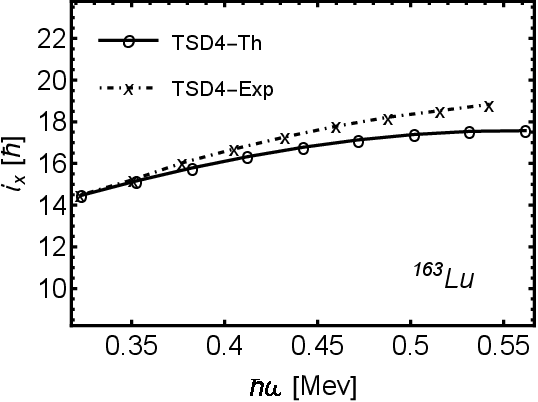}
\caption{(Results for the aligned angular momenta, $i_x$, relative to a reference $I_{ref}={\cal J}_{0}\omega+{\cal J}_{1}\omega^3$ with ${\cal J}_{0}=30\hbar^2MeV^{-1}$ and 
${\cal J}_{1}=40\hbar^4MeV^{-3}$, in $^{163}$Lu, are compared with  the corresponding experimental data \cite{Jens1,Hage}.}
\end{center}
\label{Fig.8}
\end{figure}
In Fig.12 the excitation energies normalized to a reference energy are compared with the corresponding experimental data \cite{Jens1,Hage}

\begin{figure}[ht!]
\begin{center}
\includegraphics[height=4cm,width=0.24\textwidth]{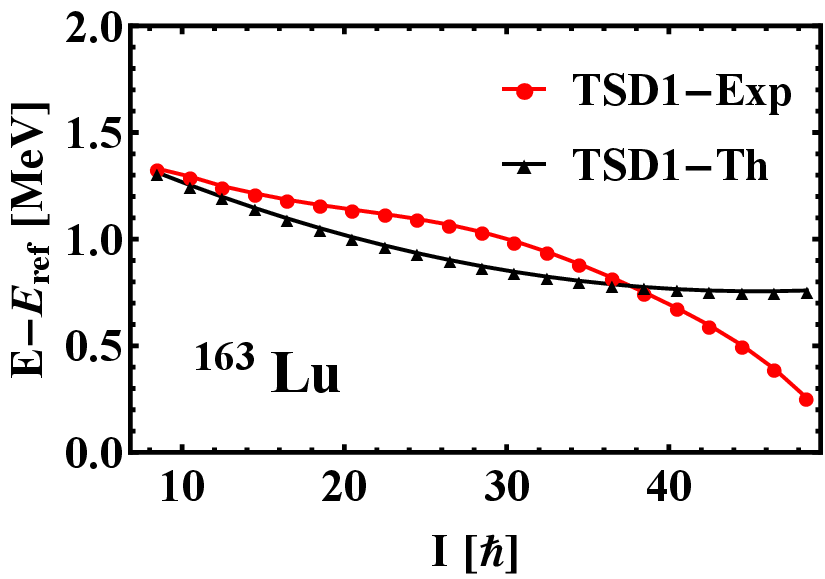}\includegraphics[height=4cm,width=0.24\textwidth]{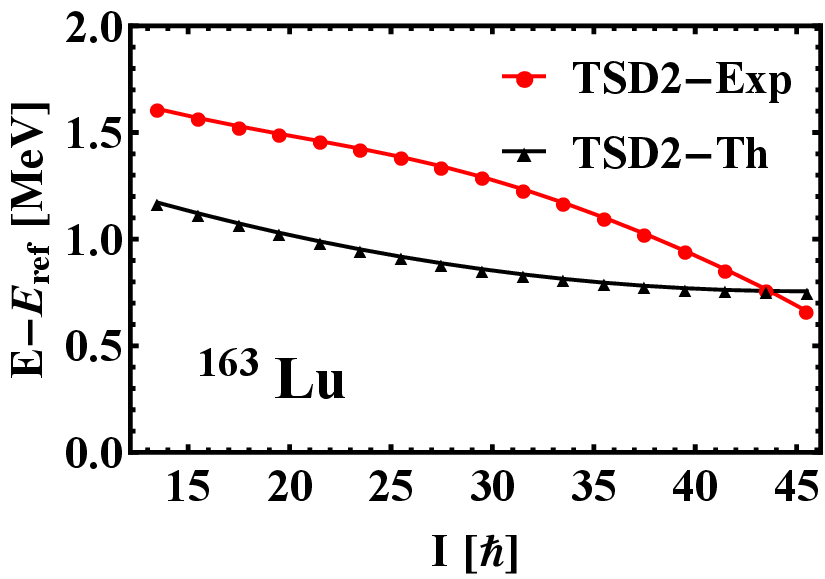}\includegraphics[height=4cm,width=0.24\textwidth]{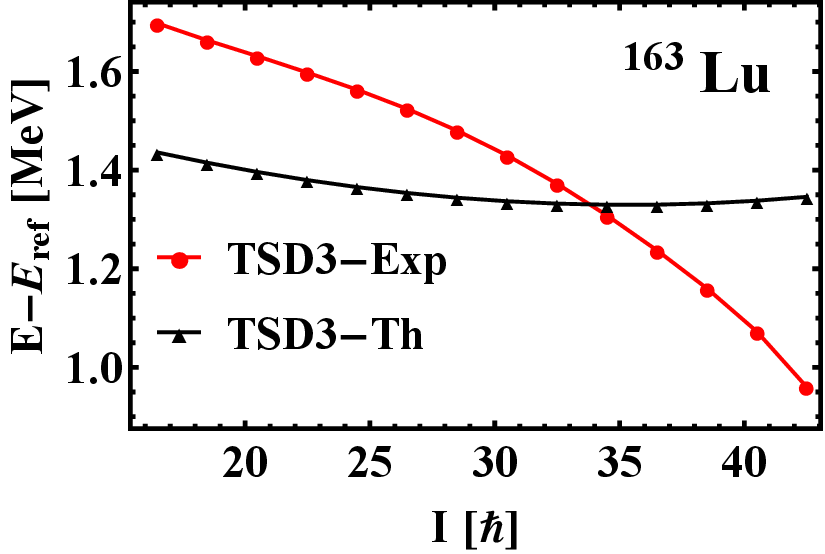}\includegraphics[height=4cm,width=0.24\textwidth]{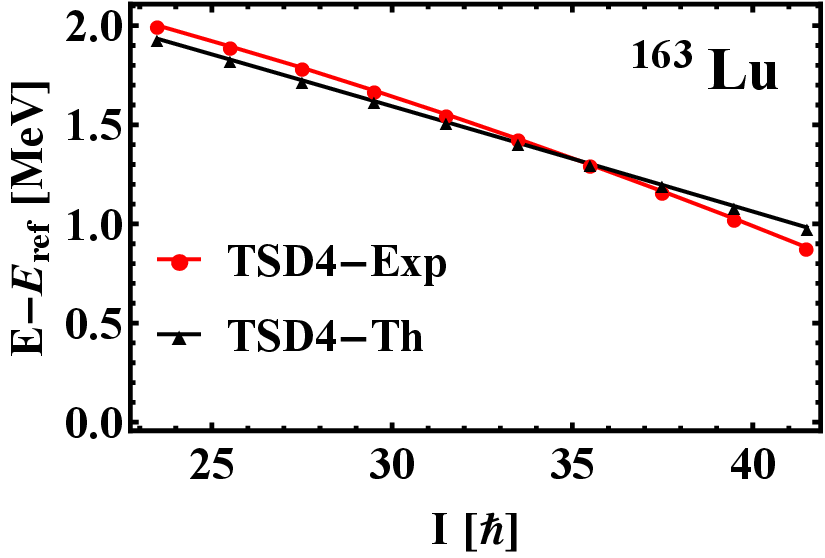}
\caption{Theoretical and experimental excitation energies for TSD1 in $^{161}$Lu normalized to the energy of a rigid rotor with an effective moment of inertia, i.e. 
$E_{REF}=0.0075I(I+1)(MeV)$, are plotted as function of the angular momentum.for the bands TSD1, TSD2, TSD3, and TSD4 of $^{163}$Lu.}
\end{center}
\label{Fig.12}
\end{figure}
The calculated and experimental dynamic moments inertia are compared in Fig. 13.
\begin{figure}[ht!]
\begin{minipage}{7cm}
\includegraphics[height=4cm,width=0.4\textwidth]{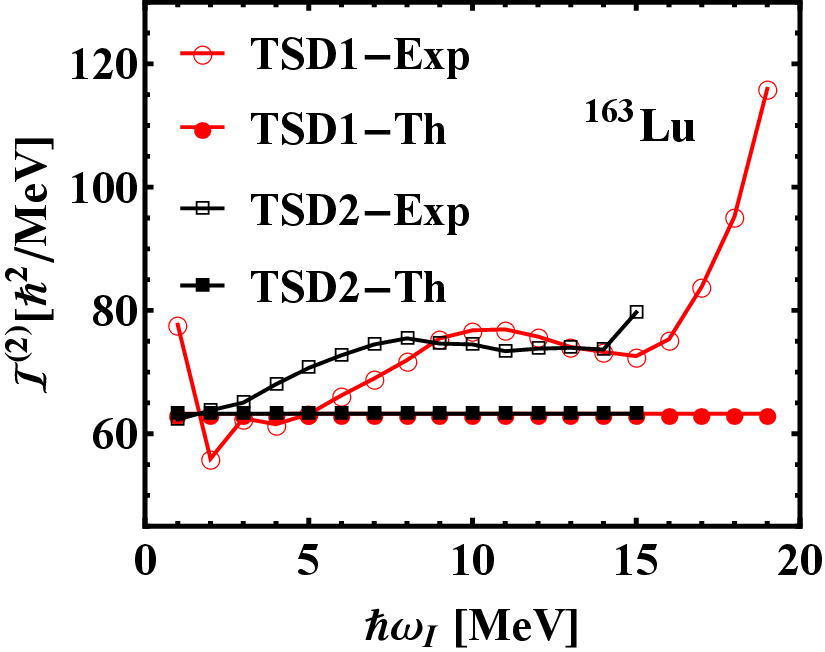}\includegraphics[height=4cm,width=0.4\textwidth]{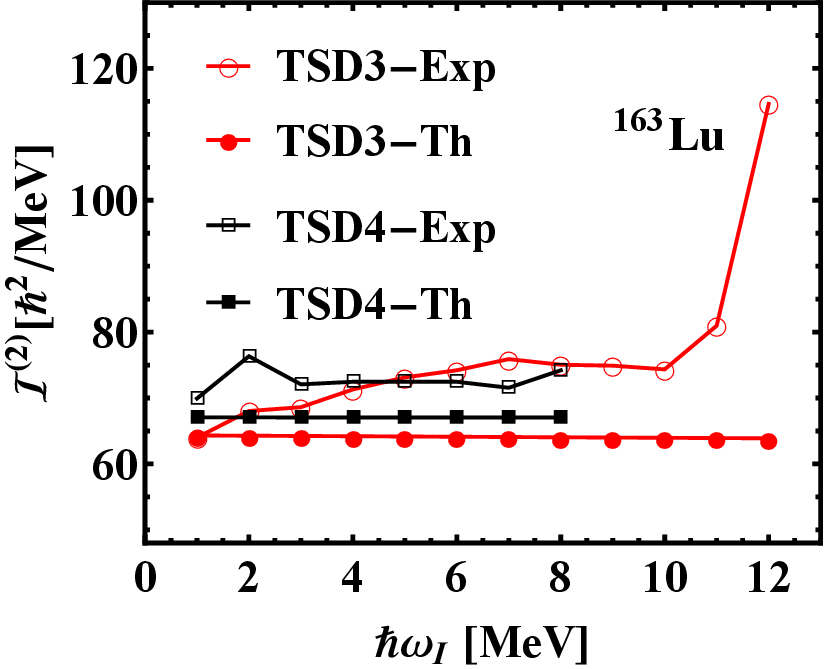}
\end{minipage}\ \ \hspace*{1.5cm}
\begin{minipage}{7cm}
\caption{(Results for the dynamic MoI's in the bands TSD1, TSD2, TSD3 and TSD4 of $^{163}$Lu, are compared with the corresponding experimental data taken from Ref.\cite{Jens1,Odeg}.}
\end{minipage}
\label{Fig.16}
\end{figure}

To calculate the quadrupole electric transition probabilities we need the expression of the wave-functions describing the states involved in the given transition and the quadrupole transition operator. The available experimental data concern the states of TSD1 and TSD2. As we saw before, the level energies from these bands account for the wobbling motion through the zero point energy. Therefore the wave function is considered to be the one corresponding to the classical minimum energy, corrected by the first order expansion term, with the coordinates deviations from the minimum quantized. Thus, one arrived at the following wave function:
\begin{equation}
\Phi^{(1)}_{IjM}={\bf N}_{Ij}\sum_{K,\Omega}C_{IK}C_{j\Omega}|IMK\rangle |j\Omega\rangle\left\{1+\frac{i}{\sqrt{2}}\left[\left(\frac{K}{I}k+\frac{I-K}{k}\right)a^{\dagger}+
\left(\frac{\Omega}{j}k'+\frac{j-\Omega}{k'}\right)b^{\dagger}\right]\right\}|0\rangle_{I},\nonumber
\label{Phi}
\end{equation}
with ${\bf N}_{Ij}$ standing for the normalization factor, and $|0\rangle_{I}$ for the vacuum state of the bosons $a^{\dagger}$ and $b^{\dagger}$ determined by the classical coordinates $\varphi$, 
$\psi$ and the corresponding conjugate momenta $r$ and $t$ through the canonical parameters $k$ and $k'$, which are analytically expressed in Appendix A of Ref.\cite{RaPoRa}.
 Expansion coefficients of the trial function corresponding to the minimum point, in terms of the normalized Wigner function, $C_{IK}$, were  analytically expressed in \cite{RaPoAl}. 

The electric quadrupole transition operator is defined by:
\begin{equation}
{\cal M}(E2,\mu)=\left[Q_0D^2_{\mu0}-Q_2(D^2_{\mu2}+D^2_{\mu-2})\right]
+e\sum_{\nu=-2}^{2}D^2_{\mu\nu}Y_{2\nu}r^2,
\end{equation}
with $Q_0$ and $Q_2$ taken as free parameters which are to be fixed by fitting two intra-band transitions for TSD1 and TSD2. 
Note that MoI's are free parameters, that is, no option for  their nature, rigid or hydrodynamic, is adopted. To be consistent with this picture, the strengths  $Q_0$ and $Q_2$ were also considered as free parameters. However, this is not consistent with the structure of the single particle potential, which considers the collective quadrupole operator as emerging from the hydrodynamic model. These are fixed by fitting the B(E2) values for one intra-band (TSD1) and one inter-band  ($TSD2\to TSD1$) transition.  The remaining B(E2) transitions and the quadrupole transition moments, listed in Tables II-V, are free of any adjustable parameter.
Results for the B(E2) values are compared with the corresponding data in Tables IV-VI.

The magnetic transition operator used in our calculations is:
\begin{equation}
 {\cal M}(M1,\mu)=\sqrt{\frac{3}{4\pi}}\mu_N\sum_{\nu=0,\pm 1}\left[g_RR_{\nu}+qg_jj_{\nu}\right]D^1_{\mu\nu},
\end{equation}
with $R_{\nu}$ denoting the components of the core's angular momentum with the corresponding gyromagnetic factor, $g_R=Z/A$, while $g_j$ is the free gyromagnetic factor for the single proton angular momentum $j(=13/2)$, which was quenched by a factor q=0.43 in order to account for the polarization effects not included in $g_j$. 
This factor takes care of the interaction of the odd-proton orbit with the currents distributed inside the core as well as of the internal structure of the proton, which may also influence its magnetic moment \cite{Iudi}.

To evaluate the transition matrix elements of the core's angular momentum, the  involved states are written in the form:
\begin{equation}
\Psi_{IM}=\frac{1}{\sqrt{2j+1}}\sum_{M_R \Omega K} C^{RjI}_{M_{R} \Omega M}C_{RK}|RM_{R}K\rangle|j\Omega\rangle .
\end{equation}
Again, the expansion coefficients of the core's wave function  in the basis of the normalized Wigner function are denoted by $C_{RK}$.
Results for the relevant B(E2) and $B(M1)$ values of the inter-band transitions $I\to I-1$ as well as for the mixing rations are collected in Table IV. Here the coordinate fluctuations around their minima are ignored since their contribution is negligible. Results branching ratios and transition quadrupole moment are collected in Tables V and VI.
{\scriptsize
\begin{table}
\begin{tabular}{|c|c|cc|cc|cc|}
\hline
&&\multicolumn{2}{c|}{$B(E2)[e^2b^2]$} &\multicolumn{2}{c|}{$B(M1)[\mu_N^2]$}&\multicolumn{2}{c|}{$\delta_{I\to(I-1)}$}\\
&&\multicolumn{2}{c|}{$I^+\to (I-1)^+$}&\multicolumn{2}{c|}{$I^+\to (I-1)^+$}&\multicolumn{2}{c|}{$[MeV.fm]$}  \\
&$I^{\pi}$&Th.  &  Exp.& Th. &Exp. &Th.&Exp.\\
\hline
&$\frac{47}{2}^+$&0.54&0.54$^{+0.13}_{-0.11}$&0.017&0.017$^{+0.006}_{-0.005}$&-1.55&-3.1$^{+0.36}_{-0.44}$\\
&$\frac{51}{2}^+$&0.49&0.54$^{+0.09}_{-0.08}$&0.018&0.017$^{+0.005}_{-0.005}$&-1.58&-3.1$\pm 0.4$$^{a)}$\\
$^{163}$Lu&$\frac{55}{2}^+$&0.44&0.70$^{+0.18}_{-0.15}$&0.019&0.024$^{+0.008}_{-0.007}$&-1.61&-3.1$\pm 0.4$$^{a)}$\\
&$\frac{59}{2}^+$&0.34&0.65$^{+0.34}_{-0.26}$&0.019&0.023$^{+0.013}_{-0.011}$&-1.64&-3.1$\pm 0.4$$^{a)}$\\
&$\frac{63}{2}^+$&0.36&0.66$^{+0.29}_{-0.24}$&0.020&0.024$^{+0.012}_{-0.010}$&-1.66&\\
\hline
\end{tabular}
\caption{\textmd{The B(E2) and B(M1) values for the transitions from TSD2 to TSD1. Mixing ratios are also mentioned. Theoretical results (Th.) are compared with the corresponding experimental (Exp.) data taken from Ref.\cite{Gorg,Jens01}. Data labeled by $^{a)}$ are from Ref.\cite{Reich}.}}
\label{Tabel 4}
\end{table}}

{\scriptsize
%\begin{widetext}
\begin{table}
\hspace*{-1cm}
\begin{tabular}{|c|c|cc|cc|c|c|cc|cc|}
\hline
&&\multicolumn{2}{c|}{$B(E2;I^+\to (I-2)^+)$} &\multicolumn{2}{c|}{$Q_I$}&&&\multicolumn{2}{c|}{$B(E2;I^+\to (I-2)^+)$} &\multicolumn{2}{c|}{$Q_I$}\\
&&\multicolumn{2}{c|}{$[e^2b^2]$}    &\multicolumn{2}{c|}{$[b]$}&&&\multicolumn{2}{c|}{$[e^2b^2]$}    &\multicolumn{2}{c|}{$[b]$}  \\
\hline
TSD1&$I^{\pi}$&Th.  &  Exp.& Th. &Exp. &TSD2&$I^{\pi}$            &   Th.       &   Exp.   &  Th.   &  Exp.\\
\hline
&$\frac{41}{2}^+$&2.80&3.45$^{+0.80}_{-0.69}$&8.89&9.93$^{+1.14}_{-0.99}$&&$\frac{47}{2}^+$&2.71&2.56$^{+0.57}_{-0.44}$&8.71&8.51$^{+0.95}_{-0.73}$\\
&$\frac{45}{2}^+$&2.74&3.07$^{+0.48}_{-0.43}$&8.77&9.34$^{+0.72}_{-0.65}$&&$\frac{51}{2}^+$&2.66&2.67$^{+0.41}_{-0.33}$&8.62&8.67$^{+0.66}_{-0.53}$\\
&$\frac{49}{2}^+$&2.69&2.45$^{+0.28}_{-0.25}$&8.66&8.32$^{+0.47}_{-0.42}$&&$\frac{55}{2}^+$&2.62&2.81$^{+0.53}_{-0.41}$&8.53&8.88$^{+0.83}_{-0.64}$\\
$^{163}$Lu&$\frac{53}{2}^+$&2.64&2.84$^{+0.24}_{-0.22}$&8.57&8.93$^{+0.38}_{-0.35}$&&$\frac{59}{2}^+$&2.58&2.19$^{+0.94}_{-0.65}$&8.46&7.82$^{+1.66}_{-1.15}$\\
&$\frac{57}{2}^+$&2.60&2.50$^{+0.32}_{-0.29}$&8.50&8.37$^{+0.54}_{-0.49}$&&$\frac{63}{2}^+$&2.54&2.25$^{+0.75}_{-0.48}$&8.39&7.91$^{+1.32}_{-0.84}$\\
&$\frac{61}{2}^+$&2.56&1.99$^{+0.26}_{-0.23}$&8.43&7.45$^{+0.49}_{-0.43}$&&$\frac{67}{2}^+$&2.51&1.60$^{+0.52}_{-0.37}$&8.34&6.66$^{+1.09}_{-0.76}$\\
&$\frac{65}{2}^+$&2.53&1.95$^{+0.44}_{-0.30}$&8.36&7.37$^{+0.82}_{-0.57}$&&$\frac{71}{2}^+$&2.49&1.61$^{+0.82}_{-0.49}$&8.28&6.68$^{+1.70}_{-1.02}$\\
&$\frac{69}{2}^+$&2.50&2.10$^{+0.80}_{-0.48}$&8.31&7.63$^{+1.46}_{-0.88}$ &&               &    &                      &         &                  \\
\hline
\end{tabular}
\caption{\textmd{The E2 intra-band transitions $I\to (I-2)$ for TSD1 and TSD2 bands. Also, the transition quadrupole moments, defined as in Ref.\cite{Hage1}, are given. Theoretical results (Th.) are compared with the corresponding experimental data (Exp.) taken from Ref. \cite{Gorg}. B(E2) values are given in units of $e^2b^2$,
while the quadrupole transition moment in $b$.}}
\label{Table 3}
\end{table}}
%\end{widetext}}
{\scriptsize
\begin{table}
%\begin{minipage}{7cm}
\begin{tabular}{|c|c|cc|cc|}
\hline
&&\multicolumn{2}{c|}{$B(E2)_{out}/B(E2)_{in}$}&\multicolumn{2}{c|}{$B(M1)/B(E2)_{in}[10^2\frac{\mu_N^2}{e^2b^2}]$} \\
&I$^{\pi}$&Th.& Exp.&Th. &Exp.\\
\hline
&$\frac{31}{2}$              & 0.29           &0.21$\pm 0.11$& &\\
&$\frac{35}{2}$              & 0.26           &0.22$\pm 0.02$&0.502           &0.439$^{+0.082}_{-0.076}$   \\
&$\frac{39}{2}$              & 0.24           &0.21$\pm 0.02$&0.560           &0.447$^{+0.077}_{-0.078}$   \\
&$\frac{43}{2}$              & 0.22           &0.22$\pm 0.02$& 0.608           &0.509$^{+0.088}_{-0.086}$   \\
$^{163}$Lu&$\frac{47}{2}$              & 0.20           &0.21$\pm 0.03$&0.650           &0.498$^{+0.091}_{-0.084}$  \\
&$\frac{51}{2}$              & 0.18           &0.21$\pm 0.02$&0.685          &0.709$^{+0.182}_{-0.196}$    \\
&$\frac{55}{2}$              & 0.17           &0.26$\pm 0.05$& & \\
&$\frac{59}{2}$              & 0.15           &0.30$\pm 0.09$& & \\
&$\frac{63}{2}$              & 0.14           &0.30$\pm 0.11$& & \\
\hline
\end{tabular}
%\end{minipage}\ \\hspace*{0.3cm}
%\begin{minipage}
\caption{\textmd{Branching ratios $B(E2)_{out}/B(E2)_{in}$ and $B(M1)/B(E2)_{in}$ of some  states from the band TSD2. Experimental data are from Refs.\cite{Hage,Scho,Amro,Jens,Jensen} and
\cite{Jens}, respectively.}}
%\end{minipage}
\label{Table 5}
\end{table}}
The alignment, the excitation energy relative to the energy of a rigid rotor with an effective moment of inertia , and the dynamic moment of inertia are plotted, in Figs 12-14, .as function of the rotatinal frequency, angular momentum and again on the rotational freuency, respectively. One notices a reasonable agreement of our results with the corresponding experimental data.
One specific feature for the wobbling motion consists of a strong $E2$ transition from the TSD2 to the TSD1 bands. This is reflected by the relative large values of the branching ratios characterizing the states from TSD2. This is confirmed by Table IV, where the calculated branching ratios are compared with the corresponding experimental data. Also, the computed ratio $B(M1)/B(E2)_{in}$ are in good agreement with the experimental data in $^{163}$Lu. Another specific wobbling feature is the large transition quadrupole moment, as shown in Table IV. From there one can see a very good agreement of our calculation results for $^{163}$Lu, and the coresponding data.
Concluding the application part of the present paper, one may say that the proposed semi-phenomenological approach seems to be an efficient tool to account for the main features of electromagnetic properties  of the even-odd $Lu$-isotopes and for the experimental data, respectively. 

 The portrait of the stationary points characterizing the energy function ${\cal H}$ split the parameter space in several regions associated to distinct nuclear phases. These are bordered by separatrices defined by the following equation:
\begin{equation}
\det \left(\frac{\partial ^2{\cal H}}{\partial (q_i)^{k}\partial (p_j)^{l}}\right)=0;\;i,j,=1,2; k,l=0,1,2;k+l=2.
\end{equation}
where the canonical conjugate coordinates are suggestively denoted as:
\begin{equation}
q_1=\varphi,\; q_2=\psi,\; p_1=r,\;p_2=t.
\end{equation}
After some algebraic manipulations,  the above equation leads to:
\begin{equation}
C=0,
\label{Ceq0}
\end{equation}
where C has the expression from Eq.(\ref{BandC}).
 Eq.(\ref{Ceq0}) splits to the following two equations:
\begin{equation}
z=f_1(x),\;\;z= f_2(x,y),
\label{f1andf2}
\end{equation}
with
\begin{eqnarray}
\hspace*{-0.4cm}f_1(x)&=&\left[\left(1-4(I-j)^2\right)x^2+\left(4I^2+4j^2-8Ij+2j+2I-2\right)x\right.\nonumber\\
&-&\left.(2I+2j-1)\right]\left[G_1\left((2I-2j-1)x-(2I-1)\right)\right]^{-1},\nonumber\\
\hspace*{-0.4cm}f_2(x,y)&=&\left[\left(1-4(I-j)^2\right)x^2+\left(1-2(I+j)\right)y^2+2\left(2(I-j)^2\right.\right.\nonumber\\
&+&\left.\left.(I+j)+1\right)xy\right]\left[G_2\left((2I-2j-1)x-(2I-1)y\right)\right]^{-1}.
\end{eqnarray}

The separatrices bordering the nuclear phases are plotted in the phase diagram given by Fig. 16.
\begin{figure}[ht!]
\begin{minipage}{7cm}
\includegraphics[height=4cm,width=0.8\textwidth]{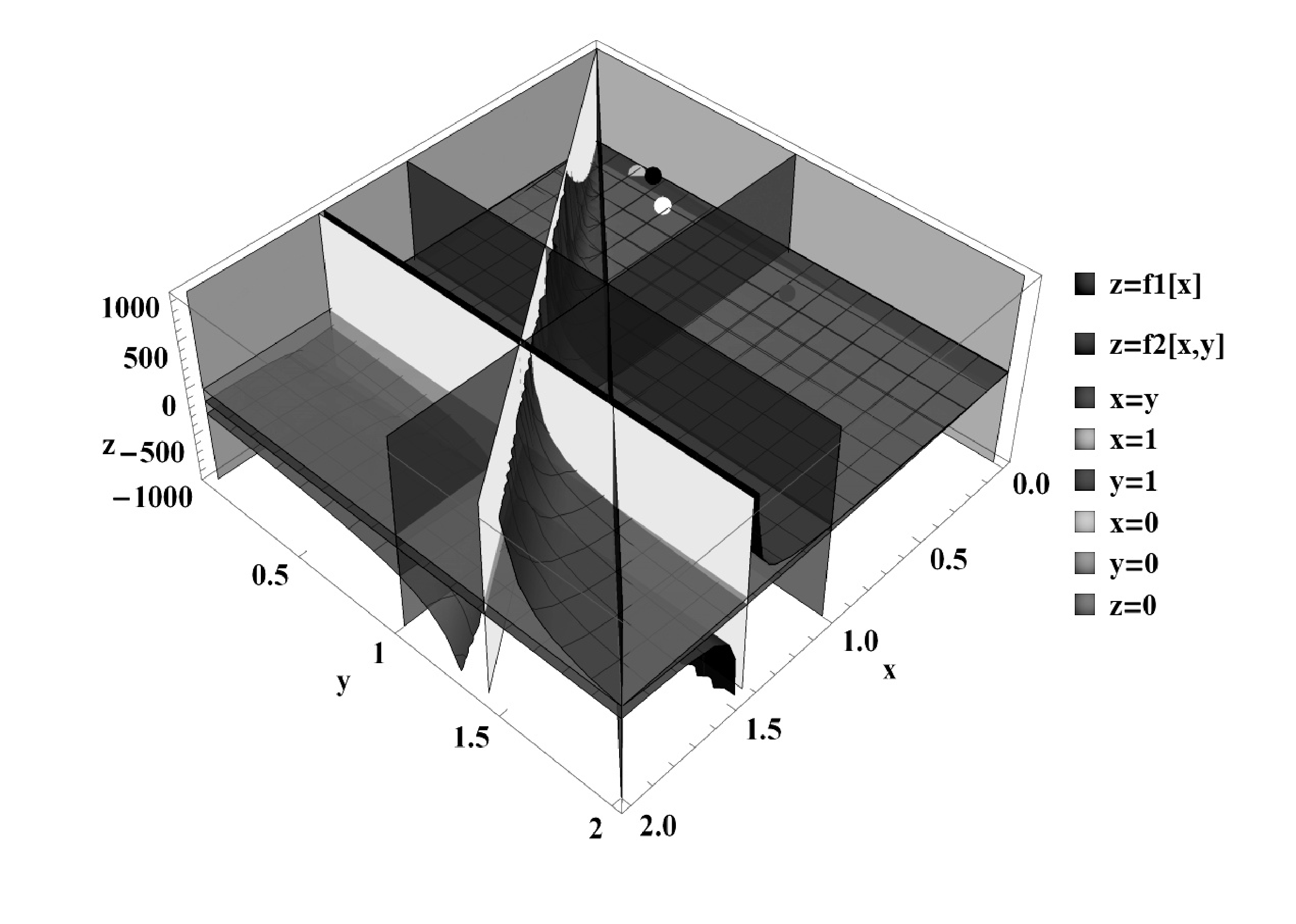}
\end{minipage} \ \ \hspace*{1cm}
\begin{minipage}{7cm}
\caption{The phase diagram for a j-particle-triaxial rotor coupling Hamiltonian with j=13/2 and I=45/2.The coordinates x,y and z are a-dimensional.}
\end{minipage}
\label{Fig.20}
\end{figure}
Here the coordinates x, y and z are defined as:
\begin{equation}
x=\frac{A_1}{A_3},\;\;y=\frac{A_2}{A_3},\;\;z=\frac{V}{A_3}.
\label{xandy}
\end{equation}
For a fixed $\gamma (=17^{o})$, the equations (\ref{f1andf2}) represent two singular surfaces, having the asymptotic planes:
\begin{equation}
x=\frac{2I-1}{2I-2j-1},\;\;y=\frac{2I-2j-1}{2I-1}x.
\label{planesxandy}
\end{equation}
%%%%%%%%%%%%%%%%%%%%%%%%%%%%%%%%%%%%%%%%%%%%%%%%%%%%%%%%%%%%%%%%%%%%%%%%%%%%%%%%%%%%%%%%%%%%%%%%%%%%%%%%%%%%%%%%%%%%
For the fixed MoI's and V shown in Table I, the coordinates $(x,y,z)$ corresponding to the isotopes $^{161,163,165,167}$Lu,respectively, are represented by small circles of different colors: purple ($^{161}$Lu)
, white ($^{163}$Lu), yellow ($^{165}$Lu) and black ($^{167}$Lu). Here only the set of MoI's corresponding to $^{163}$Lu are given. In the case of $^{161}$Lu, the coordinate y is too large and therefore drops out the range shown in Fig.16. In order to keep it inside the figure we modified y to y-7. Even so, the purple circle falls in an adjacent phase, which suggests that this isotope belongs to a different nuclear phase. Inside a given phase the classical Hamiltonian has specific stationary points. If one of these is a minimum, then the classical trajectories surround it with a certain time period. If the point in the parameter space approaches the separatrices, the period tends to infinity 
\cite{Rad98}. When $V>0$, ${\bf j}$ is always oriented along the short axis, that is the one-axis,  and the region where ${\cal I}_2>{\cal I}_1>{\cal I}_3$ is the phase where the transversal wobbling may take place. More specifically, this region is bounded by four planes, one being the diagonal plane, one is given by the second Eq. (\ref{planesxandy}), one is the plane $x=1$, and the fourth one is the plane $y=1$. There are other two planes bordering the phase of interest defined in Appendix B of Ref.\cite{Rad20}. For this region we have to depict the minimum of ${\cal H}$, if that exists. If ${\cal H}$ exhibits, indeed, a minimum in the considered sector for $\gamma$, i.e. $[0^0,60^0]$, further, the frequencies describing the small oscillations around the found minimum are to be determined.  {\it The results of our investigations \cite{Rad20} confirm the existence of a transversal mode but for  ideal restrictions \cite{Frau}, while within the Holstein-Primakoff description, the minimum for energy surface reflecting a transversal wobbling regime does not exist\cite{Tana4}, if one keeps all energy terms. Therefore, there is no contradiction between the two formalisms \cite{Frau018,Tana018}, since they deal with different Hamiltonians. Moreover, if in each of the two formalisms as well as in the present one, one keeps all terms from the starting Hamiltonian, it seems that no solution for the transversal wobbling exists.} In fact this is good example showing that an unrealistic hypothesis may lead to untrustfull/ideal results. Although in our case the transversal mode is determined by a part of the starting Hamiltonian, for the time being we cannot say whether this ideal picture is preserved when the remaining interaction is accounted for or is totally spoiled by the Coriolis interaction. An indirect answer to this question is, actually, provided by the wobbling energy behavior as function of spin, this being considered as a signature for the wobbling character. Contrary to what is stated in Ref.\cite{Frau}, in the present approach the monotony of the experimental and theoretical curves are the same, namely both are increasing functions of the angular momentum. This, in fact, confirms that the considered isotopes are longitudinal wobblers.

\renewcommand{\theequation}{8.4.\arabic{equation}}
\setcounter{equation}{0}
\section{Simultaneous description of the wobbling and chiral motion in nuclei}

Here we study an odd-mass system  consisting of an even-even core described by a triaxial rotor Hamiltonian $H_{rot}$ and a single j-shell proton moving in a quadrapole deformed mean-field:
\begin{equation}
H_{sp}=\epsilon_j+\frac{V}{j(j+1)}\left[\cos\gamma(3j_3^2-{\bf j}^2)-\sqrt{3}\sin\gamma(j_1^2-j_2^2)\right].
\label{hassp}
\end{equation}
Here $\epsilon_j$ is the single particle energy and $\gamma$, the deviation from the axial symmetric picture.
In terms of the total angular momentum ${\bf I}(={\bf R}+{\bf j}) $ and the angular momentum carried by the odd particle, ${\bf j}$, the rotor Hamiltonian is written as:
\begin{equation}
H_{rot}=\sum_{k=1,2,3}A_k(I_k-j_k)^2.
\label{hrot}
\end{equation}
where $A_k$ are half of the reciprocal moments of inertia associated to the principal axes of the inertia ellipsoid, i.e. $A_k=1/(2{\cal I}_k)$, which are considered as free parameters. 

The expressions for the single particle coupling potential, $H_{sp}$, and the triaxial rotor term, $H_{rot}$, have been previously used by many authors, the first being Davydov \cite{David}.
In the context of rigid coupling of the single particle to the core, the term $H_{sp}$ does not contribute to the equations of motion for the angular momentum components $I_k$, k=1,2,3.

We consider that the maximal moment of inertia (MoI) is ${\cal J}_1$. For a rigid coupling of the odd proton to the triaxial core, we suppose that $\bf{j}$ stays out of any principal plane: 
${\bf j}=(j\cos\theta_0, j\sin \theta_0\cos\varphi_0, j\sin \theta_0\sin\varphi_0)$.
 
We recall that within the liquid drop model (LDM) the odd nucleon may be coupled either to the deformation or to the core angular momentum. Correspondingly, the single particle angular momentum is oriented either to the symmetry axis or to the core's angular momentum \cite{Ring}. These scenarios are reached for weak and strong coupling regimes, respectively. For an intermediate coupling one may meet the situation when ${\bf j}$ lays outside the principal planes. Within a microscopic picture, the orientation of the single particle angular momentum depends on the location of the Fermi level. Thus, when the Fermi level of valence nucleon is located in the lower/upper part of a high-j
subshell, its angular momentum is oriented along the short/long axis of the triaxial core, and in the middle part
with its angular momentum easily aligned with the intermediate axis with the maximum MoI.   When the Fermi level is different from these special cases, the angular momentum of the odd proton might be oriented along a line which is different from the three mentioned axes. In these phenomenological and microscopic contests, it seems reasonable to fix the single particle angular momentum outside any principal plane of the triaxial core.

Note that the linear term in $\hat{I}$ from (\ref{hrot}) looks like the cranking term in the microscopic cranking formalism. According to the pioneering paper of Bengston \cite{Beng} the equations for a general orientation of the cranking term admit a real solution.

In this context we ask ourselves, whether the phenomenological Hamiltonian $(\ref{hrot})$ admits a harmonic solution within a classical treatment.
To this aim we dequantized the model Hamiltonian by replacing the operators $\hat{I}_k$, k=1,2,3 with the classical components of the angular momentum hereafter denoted by $x_k$,k=1,2,3, respectively, and the commutators with the Poisson brackets:
\begin{equation}
\hat{I}_k\to x_k,\;\;\;[,]\to i\{,\} .
\end{equation}
with 'i' denoting the imaginary unity and $\{,\}$ the Poisson bracket.

According to these rules, the classical energy can be written as:
\begin{eqnarray}
{\cal H}_{rot}&=&AH'+A_1I^2+\sum_{k=1,2,3}A_kj_k^2, \;\;\rm{with}\nonumber\\
H'&=&x_2^2+ux_3^2+2v_1x_1+2v_2x_2+2v_3x_3,\nonumber\\
u&=&\frac{A_3-A_1}{A_2-A_1},\; v_k=-\frac{j_kA_k}{A_2-A_1}, k=1,2,3., A=A_2-A_1.
\label{HandH'}
\end{eqnarray}
Also, the angular momentum components obey:
\begin{equation}
\{x_i,x_j\}=-\epsilon_{i,j,k}x_k.
\end{equation}
where $\epsilon_{i,j,k}$ denotes the three dimensional unity tensor.
Also, the classical counterpart of the Heisenberg quantal equation for the classical image o, of an operator $\hat{O}$,is:
\begin{equation}
\{o,H\}=i\frac{\partial o}{\partial t}.
\end{equation}
with $o$ denoting the classical image of $\hat{O}$.

Since ${\cal H}_{rot}$ and $H'$ differ from each other by one multiplicative and one additive constant, the motion described by ${\cal H}{rot}$ is readily known once that corresponding to $H'$ is given. Due to this feature, it is convenient to deal first with $H'$. Equations of motion for $H'$ are non-linear. Linearizing them one finds that the stationary points coordinates satisfy:
\begin{equation}
\frac{v_1}{x_1}-\frac{v_2}{x_2}=1;\;\;
\frac{v_1}{x_1}-\frac{v_3}{x_3}=u,\;\;
\frac{v_2}{x_2}-\frac{v_3}{x_3}=u-1.
\label{x23ofx1}
\end{equation}
From these relations we can express $x_2$ and $x_3$ in terms of $x_1$ and then insert the result in the angular momentum conservation equation:
\begin{equation}
x_1^2+x_2^2+x_3^2=I^2.
\end{equation}
It results an algebraic equation for the component $x_1$:
\begin{equation}
F(x_1)\equiv\sum_{k=0}^{6}B_kx_1^k=0,
\label{eqx1}
\end{equation}
with the coefficients $B_k$ depending on $u,v_k$ and $I$.:

We note that, since $B_0\ne 0$, the equation (\ref{eqx1}) does not admit vanishing solutions, which as a matter of fact is a specific feature for the chiral motion. The solution  for $x_1$ corroborated with Eqs.( \ref{x23ofx1}) leads to the stationary points 
$(\stackrel{\circ}{x}_1, \stackrel{\circ}{x}_2, \stackrel{\circ}{x}_3)$ for the surface of constant energy, $H'=E$.
%\end{document}
%%%%%%%%%%%%%%%%%%%%%%%%%%%%%%%%%%%%%%%%%%%%%%%%%%%%%%%%%%%%%%%%%%%%%%%%%%%%%%%%%%%%%%%%%%%%%%%%%%%%%%%%%%%%%%%%5

\subsection{Small oscillations around the deepest minimum}
The equations of motion for the components $x_k$ are non-linear. However, these can be linearized by replacing one factor of the quadratic terms with the coordinates of the deepest minimum point.

A solution of the linear system of equations may be found by searching for the linear combination:
\begin{equation}
C^*=X_1x_1+X_2x_2+X_3x_3,
\end{equation}
such that the following equation is fulfilled:
\begin{equation}
\{C^*,H'\}=\omega C^*
\end{equation}
This restriction leads to a homogeneous system of linear equations for the coefficients  $X_1,X_2,X_3$. The compatibility condition yields an equation for the frequency $\omega$:
\begin{equation}
\omega^3+3S\omega-2T=0.
\label{eqomeg}
\end{equation} 
with the coefficients given by:
\begin{eqnarray}
3S&=&-\left(2v_1-u\stackrel{\circ}{x}_1\right)\left(\stackrel{\circ}{x}_1-2v_1\right)+\left(u\stackrel{\circ}{x}_3+2v_3\right)\left(2v_3-(1-u)\stackrel{\circ}{x}_3\right)+\left(\stackrel{\circ}{x}_2+2v_2\right)
\left(2v_2+(1-u)\stackrel{\circ}{x}_2\right),\nonumber\\
2T&=&\left(2v_3+u\stackrel{\circ}{x}_3\right)\left(2v_2+(1-u)\stackrel{\circ}{x}_2\right)\left(\stackrel{\circ}{x}_1-2v_1\right)-\left(\stackrel{\circ}{x}_2+2v_2\right)\left((1-u)\stackrel{\circ}{x}_3-2v_3\right)\left(2v_1-u\stackrel{\circ}{x}_1\right).\nonumber\\
\end{eqnarray}
The solutions of Eq.(\ref{eqomeg}) are analytically given by the Cardano formula.
%%%%%%%%%%%%%%%%%%%%%%%%%%%%%%%%%%%%%%%%%%%%%%%%%%%%%%%%%%%%%%%%%%%%%%%%%%%%%%%%%%%%%%%%%%%%%%%%%%%%%%%%%%%%%%%%%%%%
%\end{document}

Note that the system under consideration exhibits two constants of motion, namely the energy and the angular momentum squared. Furthermore, there is only one independent angular momentum component; adding to this the corresponding conjugate momentum one arrives at a two dimensional phase space, which is conventionally called the {\it the reduced space}. To define this space, it is convenient to use the polar coordinates which define the two independent variables $\theta,\phi$ associated to the angular momentum ${\bf I}$. These depend, of course, on the choice for the quantization axis:

\begin{eqnarray}
&&axis\;1:\;\;x_1=I\cos\theta_1,\;x_2=I\sin\theta_1\cos\varphi_1,\;x_3=I\sin\theta_1\sin\varphi_1,\nonumber\\
&&axis\;2:\;\;x_2=I\cos\theta_2,‌\;x_3=I\sin\theta_2\cos\varphi_2,\;x_1=I\sin\theta_2\sin\varphi_2,\nonumber\\
&&axis\;3:\;\;x_3=I\cos\theta_3,\;x_1=I\sin\theta_3\cos\varphi_3,\;x_1=I\sin\theta_3\sin\varphi_3.
\end{eqnarray}
For each of the above choices, the Hamiltonian $H'$ becomes a function of the conjugate  coordinates $(x_k,\phi_k)$ which admits a minimum. Expanding $H'$ in the first order around the minimum point. The linear equations in the deviations describe a harmonic motion of frequencies:
{\scriptsize
\begin{eqnarray}
\omega^{(1)}&=&2\left[\left(\cos^2\stackrel{\circ}{\varphi}_1+u\sin^2\stackrel{\circ}{\varphi}_1+\frac{1}{I}(v_2\cos\stackrel{\circ}{\varphi}_1+v_3\sin\stackrel{\circ}{\varphi}_1)\right)\left((I^2-\stackrel{\circ}{x}_1^2)(1-u)
\cos2\stackrel{\circ}{\varphi}_1+\frac{1}{I}(I^2-\frac{\stackrel{\circ}{x}_1^2}{2})(v_2\cos\stackrel{\circ}{\varphi}_1+v_3\sin\stackrel{\circ}{\varphi}_1\right)\right]^{1/2},\nonumber\\
\omega^{(2)}&=&2\left[\left(1-u\cos^2\stackrel{\circ}{\varphi}_2-\frac{1}{I}\left(v_1\sin\stackrel{\circ}{\varphi}_2+v_3\cos\stackrel{\circ}{\varphi}_2\right)\right)
\left((\stackrel{\circ}{x}_2^2-I^2)u\cos2\stackrel{\circ}{\varphi}_2+(\frac{\stackrel{\circ}{x}_2^2}{2I}-I)\left(v_1\sin\stackrel{\circ}{\varphi}_2+v_3\cos\stackrel{\circ}{\varphi}_2
\right)\right)\right]^{/2},\nonumber\\
\omega^{(3)}&=&2\left[\left(-\sin^2\stackrel{\circ}{\varphi}_3+u-\frac{1}{I}(v_1\cos\stackrel{\circ}{\varphi}_3+v_2\sin\stackrel{\circ}{\varphi}_3))\right)
\left((I^2-\stackrel{\circ}{x}_3^2)\cos2\stackrel{\circ}{\varphi}_3+(\frac{\stackrel{\circ}{x}_3^2}{2I}-I)(v_1\cos\stackrel{\circ}{\varphi}_3+v_2\sin\stackrel{\circ}{\varphi}_3)\right)\right]^{1/2}.
\end{eqnarray}}
%\end{document}
If these frequencies are all real, then they describe the wobbling frequencies for the motion along the axes 1,2,3, respectively.
It is interesting to see what is the relations between the   frequencies given above and the solutions of the cubic equation (\ref{eqomeg}). This issue will be pointed out in what follows.

In the space of angular momentum, a chiral transformation is equivalent to the space inversion operation, i.e. $C={\bf I}\to -{\bf I}$. Due to the linear terms in angular momentum components, the Hamiltonian $H'$ is not invariant to chiral transformations. On the other hand, if there exists an operator $O$ which satisfies the relation,
\begin{equation}
\{H,O\}=0,
\label{Hanticom}
\end{equation}
then, if $\Psi$ is an eigenfunction of $H$ corresponding to the eigenvalue $\lambda$, it results that $O\Psi$ is also an eigenfunction of $H$ with the eigenvalue $-\lambda$.
Therefore, the eigenvalues $\lambda$ and -$\lambda$ are mirror images of one another.
In our case $H'$ is a sum of two terms, one invariant, $H_1$, and another non-invariant, $H_2$, to chiral transformations C. The non-invariant term $H_2$ and the transformation C satisfy 
Eq.(\ref{Hanticom}). Due to this feature the eigenvalues of $H'$ are mirror images of those for $CH'C^{-1}$.The two sets of energies define the so called chiral bands.
We note that $CH'C^{-1}$ is obtained from $H'$ by changing $v_k\to  -v_k$, which results that the wobbling frequencies, $\omega^{(k)}_{ch}$, built up with $CH'C^{-1}$ are obtained from those obtained with $H'$ with  the transformation $v_k\to  -v_k$.

The notations $H'^{(k)}_{min}$ and $H'^{(k)}_{ch,min}$ are used for minimal energy when the axis "k" is the quantization axis. 
 The energies, defined with the above mentioned frequencies are angular momentum dependent. For a given $n$ and I=$\alpha$+2n with $\alpha$  being the signature, the set of energies $E^{(k)}_{I,n}$ defines a wobbling band, while
$E^{(k)}_{ch,I,n}$ the chiral partner band. In this way we found out a set of states which are simultaneously of wobbling and chiral character.
The wobbling motion is conciliated with the ingredient that the rotation axis is outside the principal planes.

%\end{document}

Here we present an illustrative example. Thus, we consider an odd particle from the single particle orbit $j=i_{13/2}\hbar$ moving around a triaxial rigid rotor core with the moments of inertia (MoI):
\begin{equation}
({\cal J}_1, {\cal J}_2, {\cal J}_3)= (60, 20, 30)\hbar^2MeV^{-1}.
\end{equation}
The composite system moves in a state of angular momentum $I=35/2\hbar$. The odd particle is rigidly coupled to the core : ${\bf j}$ are: ${\bf j}=(j,\theta_0,\varphi_0)=(13/2,\pi/4,\pi/4)$. 
The stationary points for the equations of motion for the classical angular momentum components $x_k$, k=1,2,3, obey a set of equations which leads to an algebraic sixth-order equation for $x_1$
,i.e $F(x_1,I)=0$. 
This equation admits four real solutions for $x_1$:$-13.062; -8.811; -1.81; 16.185 [\hbar]$. Making use of relations expressing $x_2$ and $x_3$ in terms of $x_1$ one arrives at the final result:
\begin{eqnarray}
(\stackrel{\circ}{x}_1, \stackrel{\circ}{x}_2, \stackrel{\circ}{x}_3)=\left(\begin{matrix} & -13.062&   5.916&  10.029\\
                                   & -8.811 &  6.595&  13.588\\
                                   & -1.810& 16.886 &  -4.223\\
                                    & 16.185&   4.269&   5.062\end{matrix}  \right)\hbar.
\label{stavect}
\end{eqnarray}

To the four stationary vectors  the following  classical energies correspond:
\begin{eqnarray}
H_{rot}=\left(\begin{matrix}&axis-1&axis-2&axis-3\\
                            &3.542&3.027&2.887 \\
                            &3.559&3.093&2.839\\
                            &5.921&4.920&5.793\\
                            &1.200&1.452&1.424\end{matrix}\right) \rm{MeV}.
\label{energ}
\end{eqnarray}
For example, for the stationary angular momenta components from the row 1 of Eq. (\ref{stavect}), the energies of the row 1 from Eq.(\ref{energ}) correspond, for the situations when the quantization axes are the axis-1,axis-2 and axis-3, respectively.

The frequencies characterizing the linearized equations of motion satisfy a third order algebraic equation. The minimum value for the energy $H_{rot}$ when ${\bf I}\parallel {\bf j}$ ,
i.e. when the two angular momenta are aligned, is 1.765 MeV. 
 With this data there exists a real solution for the wobbling frequency:
\begin{equation}
\omega = 0.362 {\rm MeV.}
\end{equation}
The chiral partner state has the frequency equal to 3.651 MeV.

The coordinates  and spins of all minima are collected in Table VII; these minima  are taken from the contour plots shown in Figs. 17, respectively.
\begin{table}[h!]
\begin{tabular}{|c|cc| c c c|c|}
\hline\\
quantization axis&$\theta_{min}$[rad]&$\varphi_{min}$[rad]&$I_1[\hbar]$&$I_2[\hbar]$&$I_3[\hbar]$&$H_{rot,min}$[MeV]\\
\hline
axis-1          &0.388 &0.8703&16.198& 4.269& 5.063&1.203\\
axis-2          &1.206 &1.236&15.443 &6.238& 5.370&1.381\\
axis-2          &1.104 &-0.983&- 13.003& 7.879& 8.666&2.960\\
axis-3&           1.124    &0.283  &15.152&4.403&7.569&  1.361       \\
\hline
\end{tabular}
\caption{\textmd{Coordinates of the minima points for $H_{rot}$ and the corresponding values of the spin components.}} 
\end{table}
Furthermore, we studied the equations of motion for $H'$ in the reduced space of generalized phase space coordinates.
\begin{figure}[h!]
\begin{center}
%\hspace*{-2cm}
\includegraphics[height=4cm,width=0.35\textwidth]{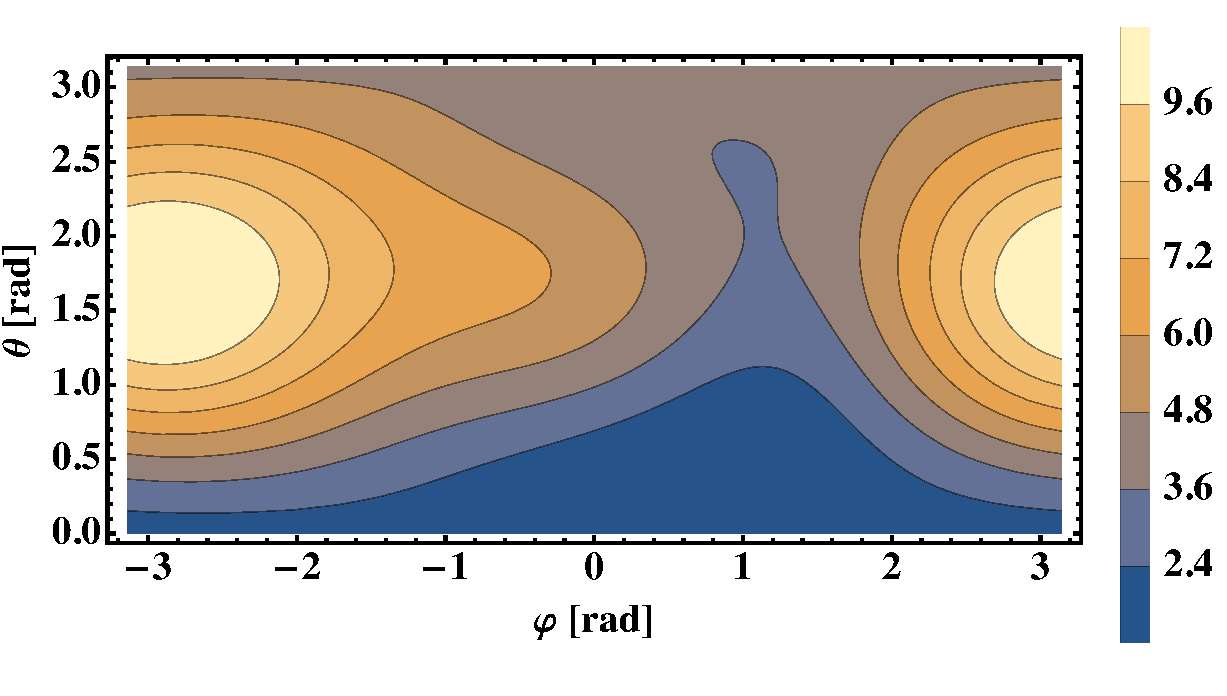}\includegraphics[height=4cm,width=0.35\textwidth]{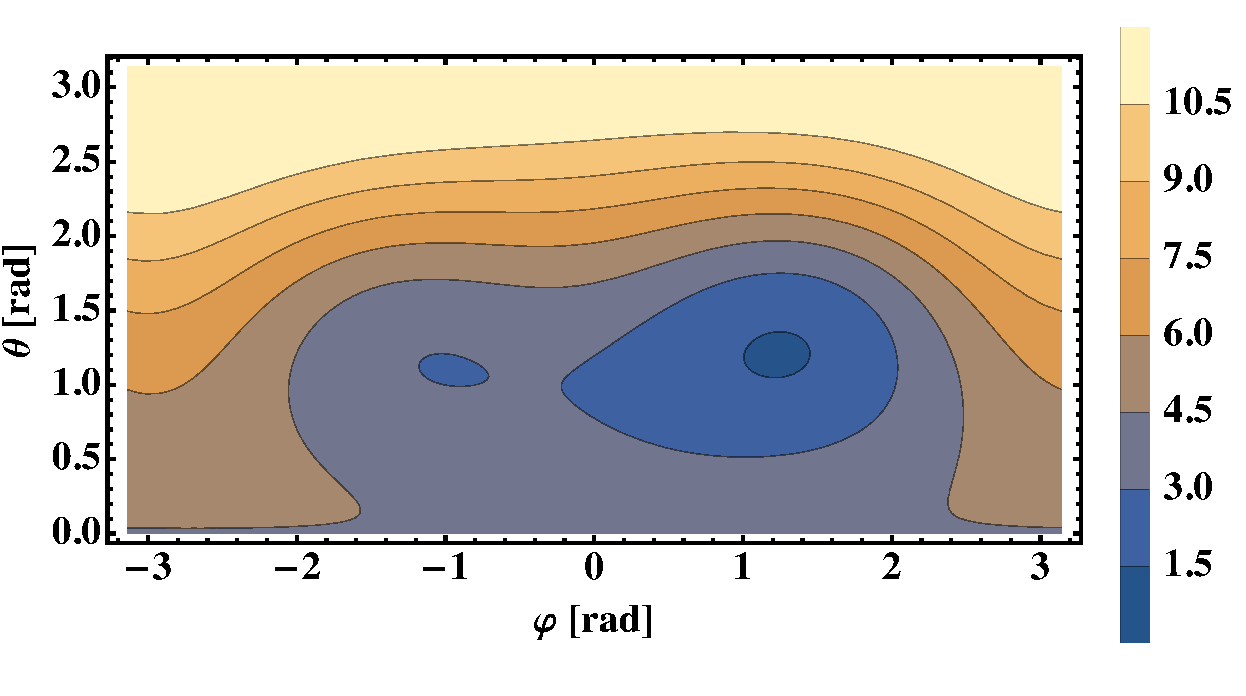}\includegraphics[height=4cm,width=0.35\textwidth]{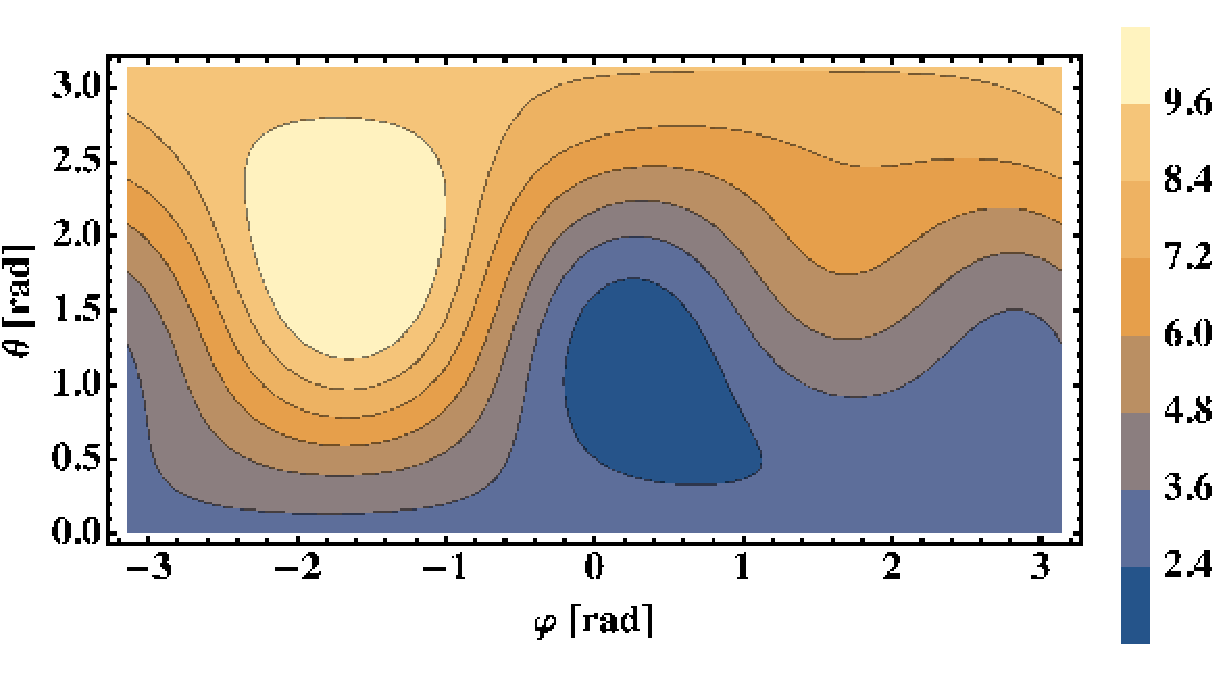}
\end{center}
\caption{  The energy function is $H_{rot}$ given by Eq. (2.2)
 as a function of the polar coordinates. Contour plot when the axis-1, axis-2 and axis-3 are the quantization axes, respectively.}
\label{Fig.2}
\end{figure}

The coordinates  and spins of all minima are collected in Table I
%\clearpage
\begin{figure}[h!]
\begin{center}
%\vspace*{-2cm}
\includegraphics[height=4cm,width=0.35\textwidth]{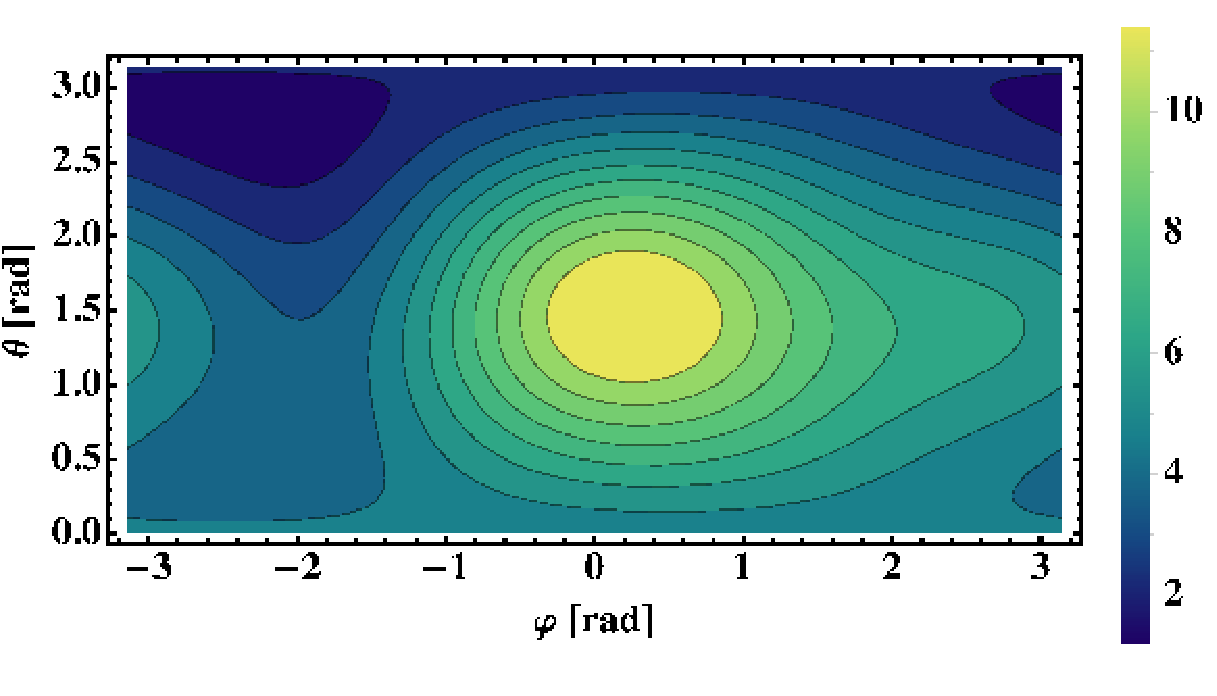}\includegraphics[height=4cm,width=0.35\textwidth]{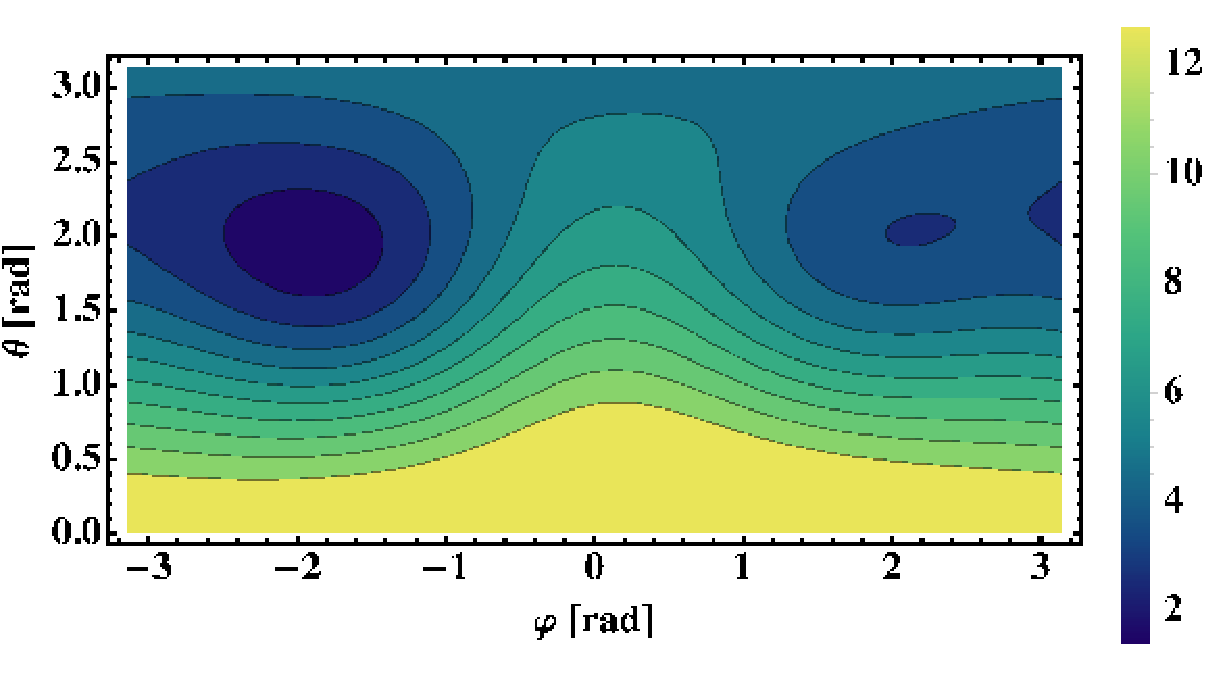}\includegraphics[height=4cm,width=0.35\textwidth]{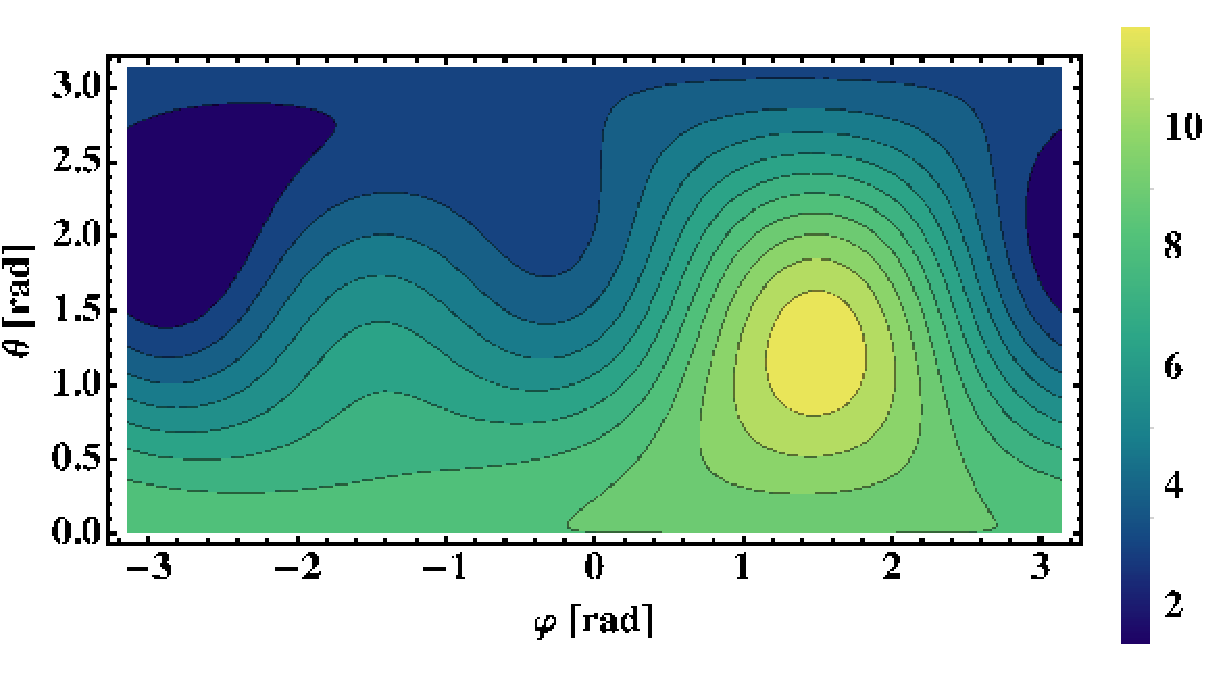}
\end{center}
\caption{ Contour plot when the axis-1,axis-2 and axis-3 are the quantization axes, respectively. The energy function is $H^{ch}_{rot}$, the chiral image of the classical energy.}
\label{Fig.6}
\end{figure}

%\end{document}

A major conclusion of this analysis is that irrespective of the chosen quantization axis, the deepest minimum of $H_{rot}$ is met for an angular momentum lying outside any principal plane of the inertia ellipsoid which is a prerequisite of a chiral motion.
\begin{center}
\begin{table}[h!]
\begin{tabular}{|c|cc| c c c|c|}
\hline\\
quantization axis&$\theta_{min}$&$\varphi_{min}$&$I_1[\hbar$&$I_2[\hbar]$&$I_3[\hbar]$&$H^{ch}_{rot,min}[MeV]$\\
\hline
axis-1          &2.753 &- 2.27 &- 16.198& - 4.269& - 5.063&1.202 \\
axis-1          &2.894 & 3.141 &- 16.965 &- 4.293 &$\approx$0.0&1.478 \\
axis-2          &1.935& - 1.905&- 15.443& - 6.238& - 5.370&1.381 \\
axis-2          & 2.148&3.141&$\approx$ 0& - 9.553 &- 14.662&2.873 \\
axis-3          &2.018& - 2.859& - 15.152& - 4.403& - 7.568&1.361 \\
axis-3          &2.021& 3.142 &- 15.754& $\approx$ 0& - 7.620&1.719 \\
\hline
\end{tabular}
\caption{{Coordinates of the minima points for the chirally transformed Hamiltonian, $H^{ch}_{rot}$, and the corresponding values of the spin components}} 
\end{table}
\end{center}
%\clearpage

%\clearpage
The transformed Hamiltonian $H^{ch}_{rot}$ has the contour plots graphically represented in Figs. 18 for the quantization axes 1, 2 and 3 respectively, with the minima coordinates collected in Table VIII.

The wobbling frequencies corresponding to the deepest minima when the quantization axis is the axis 1, 2 and 3 respectively, have the values:0.245, 0.209, 0.210 MeV, respectively.
One notes that when axis-1 is the quantization axis, the Hamiltonians $H_{rot}$ and $H^{ch}_{rot}$ have the same wobbling frequencies,  while for axis-2 as quantization axis the wobbling frequencies are very close to each other. Moreover the two Hamiltonians have the same/almost the same values in the respective minima. This is a reflection of the fact that for the two cases the chiral invariant part of $H_{rot}$ prevails over the non-invariant part.
Consequently, for the cases when the quantization axis is the axis-1 or 2, one can build up two wobbling bands with a similar structure which results in having a twin pair of bands.

 Note that contrary to the previous publications, where the odd nucleon was  rigidly fixed either to an axis \cite{Bu} or to a principal plane \cite{Rad2020} of the inertia ellipsoid, here the rigid coupling is achieved to a direction which does not belong to a principal plane. In this way one conciliates between the two signatures of triaxial nuclei, the wobbling and chiral motion, these being simultaneously considered.

\renewcommand{\theequation}{8.5.\arabic{equation}}
\setcounter{equation}{0}
\section{Boson description of the wobbling motion}
Assuming a rigid coupling of an odd nucleon to a triaxial core, the Hamiltonian for the even-odd system may be approximated as:
\begin{equation}
\hat{H}_{rot}=\sum_{k=1,2,3}A_k(\hat{I}_k-\hat{j}_k)^2,
\end{equation}
with $A_k=\frac{1}{2{\cal J}_k}$  and $I$ standing for the total angular momentum.

For a rigid coupling of the odd proton to the triaxial core, we suppose that $\bf{j}$ stays in the principal plane (1,2).
Also,  we consider that the maximal moment of inertia (MoI) is ${\cal J}_2$; furthermore, we expand the linear term in the first order of approximation:
\begin{equation}
\hat{I}_2=I\left(1-\frac{1}{2}\frac{\hat{I}_1^2+\hat{I}_3^2}{I^2}\right).
\end{equation}
Thus, the Hamiltonian acquires the form:
\begin{equation}
\hat{H}_{rot}=A\hat{H}^{\prime}+(A_1I^2-A_2j_2I)+\sum_{k=1,2}A_k\hat{j}_k^2,
\label{hahhasprim}
\end{equation}
where the following notations have been used:
\begin{eqnarray}
\hat{H}^{\prime}&=&\hat{I}_2^2+u\hat{I}_3^2+2v_0\hat{I}_1,\;\;\rm{with}\nonumber\\
A&=&A_2(1-\frac{j_2}{I})-A_1,\;\;u=\frac{A_3-A_1}{A},\;\;v_0=\frac{-A_1j_1}{A}.
\end{eqnarray}
For what follows, we suppose that the MoI's are such that $1>u>0$.

 Note that $\hat{H}^{\prime}$ looks like a Hamiltonian for a triaxial rotor amended with a term, which cranks the system to rotate around the one-axis. It is convenient to choose  the cranking axis as quantization axis. Moreover, it is useful to express the considered Hamiltonian in terms of the raising and lowering angular momenta operators:
\begin{equation}
\hat{I}_{\pm}=\hat{I}_2\pm i\hat{I}_3,\;\;\hat{I}_0=\hat {I}_1.
\end {equation}
In the intrinsic frame of reference, the angular momentum components satisfy the commutation relations:
\begin{equation}
\left[\hat{I}_{-},\hat{I}_{+}\right]=2\hat{I}_{0},\;\;\left[\hat{I}_{\mp},\hat{I}_{0}\right]=\mp\hat{I}_{\mp}.
\end{equation}
In terms of the new variables, one obtains:
\begin{equation}
\hat{H}^{\prime}=\frac{1-u}{4}\left(\hat{I}_{+}^{2}+\hat{I}_{-}^{2}\right)+\frac{1+u}{4}\left(\hat{I}_{+}\hat{I}_{-}+\hat{I}_{-}\hat{I}_{+}\right)+2v_0\hat{I}_0.
\end{equation}
The Schr\"{o}dinger equation associated to $\hat {H}^{\prime}$,
\begin{equation} 
\hat {H}^{\prime}|\Psi\rangle =E|\Psi\rangle,
\end{equation}
is further written in terms of the conjugate variables $q$ and $\frac{d}{dq}$, by using the following representation for the angular momentum components:

\begin{equation}
\hat{I}_{\mp}=i\frac{c\pm d}{k's}\left(I\mp \hat{I}_0\right),\;\;\;\hat{I}_0=Icd-s\frac{d}{dq}\equiv \hat{I}_1,
\label{amqdq}
\end{equation} 
where $s, c$ and $d$ denote the Jacobi elliptic functions:
\begin{eqnarray}
s&=&sn(q,k),\;\;c=cn(q,k),\;\;d=dn(q,k),\;\;\rm{with}\nonumber\\
k&=&\sqrt{u},\;\;k^{\prime}=\sqrt{1-k^2},\;\;q=\int_{0}^{\varphi}\left(1-k^2\sin^2(t)\right)^{-1/2}dt\equiv F(\varphi,k).
\end{eqnarray}
Their connection with the trigonometric function is given by:
\begin{equation}
s=\sin\varphi,\;\;c=\cos \varphi,\;\;d=\sqrt{1-k^2s^2}.
\end{equation}

The functions $s, c$ and $d$ are periodic  in $\varphi$, with the periods
$4K, 4K$ and $2K$ respectively, where:
\begin{equation}
K=F(\frac{\i}{2},k)=\frac{\pi}{2} {_2}F{_1}(\frac{1}{2},\frac{1}{2},1;k^2).
\label{Ka}
\end{equation}
The standard notation for the hyper-geometric function $_2F_1(\alpha,\beta,\gamma;\epsilon)$, has been used. 
It is worth mentioning that the transformation (\ref{amqdq}) preserves the commutation relations obeyed by the angular momentum components (2.6).

In terms of the newly introduced conjugate coordinates, $\hat{H}^{\prime}$ becomes:
\begin{equation}
    \hat{H}^{\prime}=-\frac{d^2}{dq^2}-2v_0s\frac{d}{dq}+I(I+1)s^2k^2+2v_0cdI.
    \label{Hqdq}
    \end{equation}
Changing the wave-function by the transformation:
\begin{equation}
|\Psi\rangle=(d-kc)^{-\frac{v_0}{k}}|\Phi\rangle,
\end{equation}
the Schr\"{o}dinger equation acquires a new form, where the kinetic and potential energies are separated:
\begin{equation}
\left[-\frac{d^2}{dq^2}+V(q)\right]|\Phi\rangle =E|\Phi \rangle .
\label{Scheq}
\end{equation}
The potential energy term has the expression:
\begin{equation}
V(q)=\left[I(I+1)k^2+v_0^2\right]s^2+(2I+1)v_0cd.
\end{equation}

\begin{figure}
%\hspace*{-1.5cm}
\begin{center}
\includegraphics[height=4cm,width=0.35\textwidth]{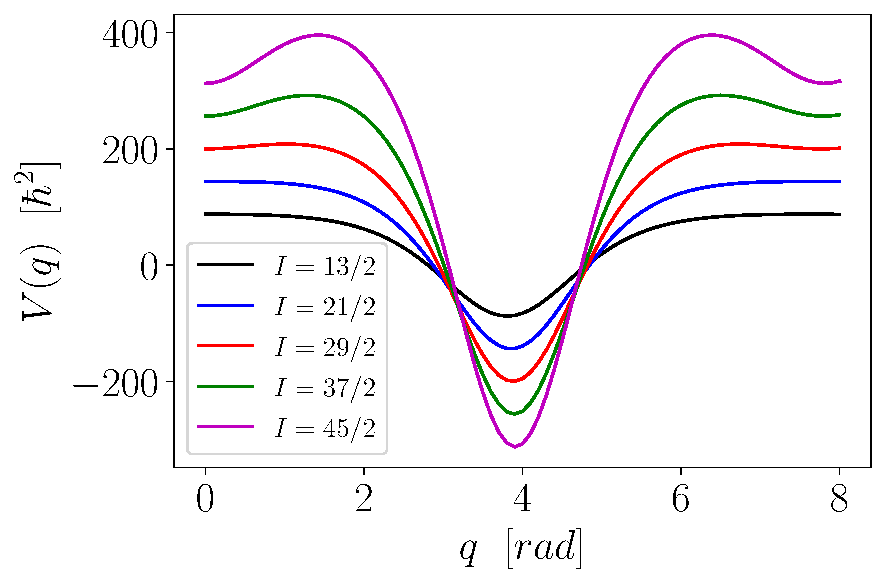}\includegraphics[height=4cm,width=0.35\textwidth]{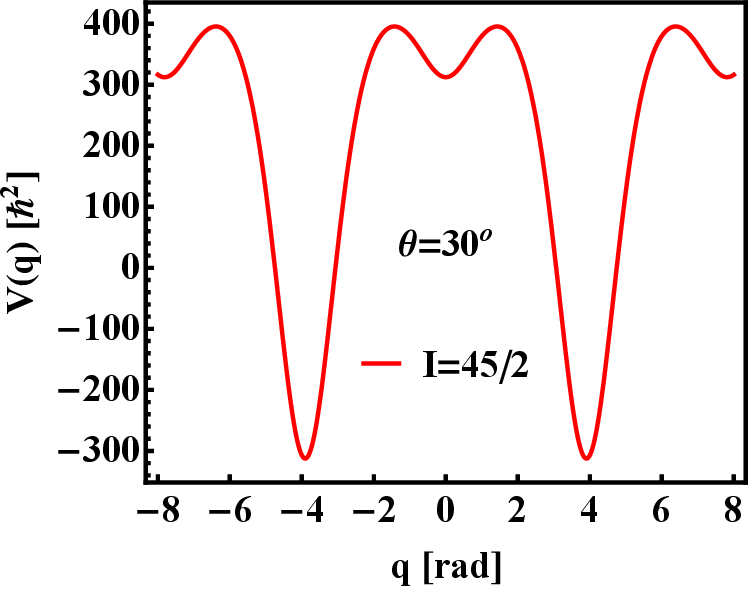}\includegraphics[height=4cm,width=0.35\textwidth]{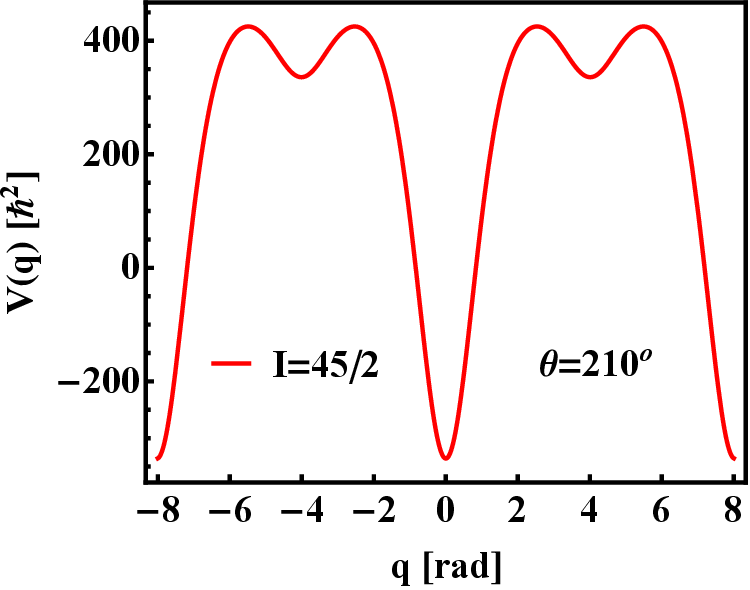}
\caption{ The potential energy is plotted as function of $q$ for a particular set values for the moment of inertia ($MoI$) 
: ${\cal J}_2:{\cal J}_3:{\cal J}_1=100:40:20\hbar^2 MeV^{-1}$,the odd particle angular momentum j=13/2 and $\theta=\pi/6$. The middle and the right figures correspond to $\theta = \pi/6 $ and $\theta=7\pi/7$, respectively. In both cases the total angular momentum is I=45/2.}
\end{center}
\label{Fig.3}
\end{figure}

 Note that V(q) is invariant with respect to the transformation $q\to-q$. This leads to the fact that in the interval, for example of
 [-4K,4K], the potential exhibits two deep symmetric wells with  degenerate minima, and three local minima  in $q=0, \pm 4K$. States inside the local minima are meta-stable since they are tunnelling to the adjacent deep minima. The states in the deepest wells are degenerate.
%%%%%%%%%%%%%%%%%%%%%%%%%%%%%%%%%%%%%%%%%%%%%%%%%%%%%%%%%%%%%%%%%%%%%%%%%%%%%%%%%%%%%%%%%%%%%%%%%%%%%%%%%%%
The shape of the potential V(q) in the interval of [0,4K] is shown in Fig.19, for a few angular momenta I. To visualize the symmetry mentioned above, we plotted V(q) in a larger interval, namely [-4K,4K]. Denoting by $\psi_+$ the wave function of a state in the right deepest well, and by $\psi_-$ the function corresponding to the same energy as the former state, but the left deepest minimum, they  are both spread over the whole interval of [-4K,+4K]. However, the sum  $\psi_++\psi_-$ is mainly located in the right deepest well, while the difference $\psi_+-\psi_-$ is mainly spread inside the left deepest well.

The formalism described above was applied to $^{135}$Pr. The excitation energies in three bands, conventionally called band 1 (B1), band 2 (B2), and band 3 (B3), and the  electromagnetic properties of the states have been described by a simple Hamiltonian (5.1), associated to the even-even core and the odd proton, which stays in the orbital $h_{11/2}$.The core properties are
simulated by a triaxial core with the moments of inertia ${\cal J}_k$ (k=1,2,3), considered to be free parameters,  while the odd proton is rigidly coupled to the core and having the angular momentum $j=11/2$,
placed in the inertial plane (1,2), and having the polar angle $\theta$. Thus, the approach involves four free parameters ${\cal J}_k$  (k=1,2,3), and $\theta$, which were fixed by a least mean square procedure, fitting the excitation energies for the three bands.
{\scriptsize
\begin{table}[h!]
\begin{tabular}{|c|c|c|c|c|c|}
\hline
 ${\cal I}_1$& ${\cal I}_2$ &${\cal I}_3$& $\theta$ &nr. of&r.m.s.\\
$[\hbar^2/MeV]$           & $[\hbar^2/MeV]$&$[\hbar^2/MeV]$&  [degrees] & states& [MeV]              \\
\hline
89                        &   12                      &      48                   & -71 &20       &0.174\\
\hline
\end{tabular}
\caption{\textmd{The MoI's, and the parameter $\theta$ as provided by the adopted fitting procedure. }}
\label{Table 1}
\end{table}}
The parameters yielded by the fitting procedure are listed in Table 9. With the parameters thus determined, the excitation energies are readily obtained:
\begin{eqnarray}
E^{exc;1}_{I} &=& A_1 I^ 2 +(2I+1)A_1 j_{1}-IA_2 j_2 + \omega_{I}/2-E_{11/2},\;\; I=R+j,\;\;R=0,2,4,...,\nonumber\\
E^{exc;2}_{I} &=& A_1 I^ 2 +(2I+1)A_1 j_{1}-IA_2 j_2 + \omega_{I}/2-E_{11/2},\;\; I=R+j,\;\;R=1,3,5,...,\nonumber\\
E^{exc;3}_{I+1} &=& A_1 I^ 2 +(2I+1)A_1 j_{1}-IA_2 j_2 + 3\omega_{I}/2-E_{11/2},\nonumber\;\;I=R+j,\;\;R=1,3,5,....
\label{exen}
\end{eqnarray}
 They are visualized in Figs.20 and compared with the corresponding experimental data taken from Refs.\cite{Matta,Sen}. From there, one sees the quality of the agreement with the data, that might be appraised by the r.m.s. of the deviation which is also given in Table 9. We may conclude that the agreement between theoretical, and experimental results is good.

Note that in the present approach there exists an additional wobbling frequency characterizing the bands 2 and three and determined variationally for the signature unfavoured states.
In this case the standard definition chosen for the experimental wobbling is not appropriate.
\begin{figure}[h!]
\begin{center}
\includegraphics[height=4cm,width=0.35\textwidth]{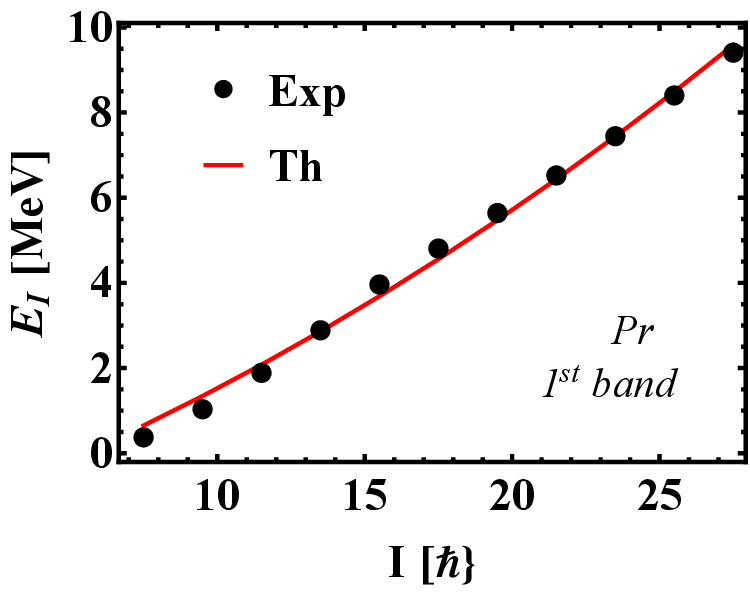}\includegraphics[height=4cm,width=0.35\textwidth]{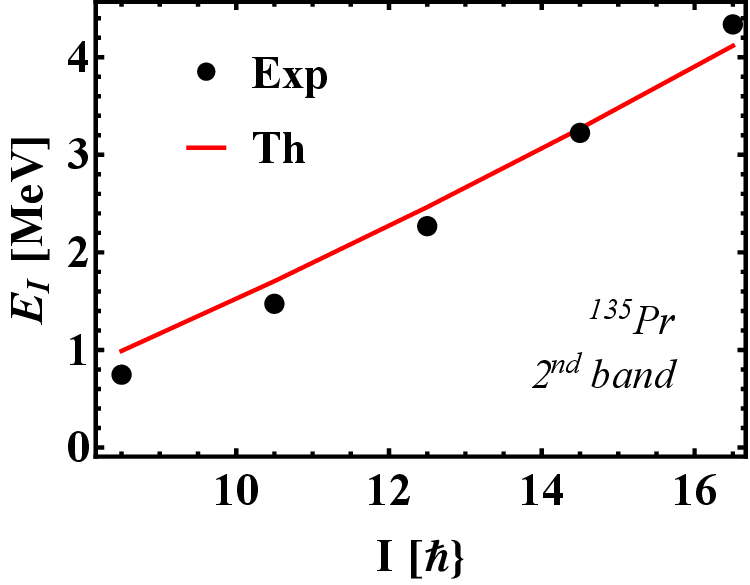}\includegraphics[height=4cm,width=0.35\textwidth]{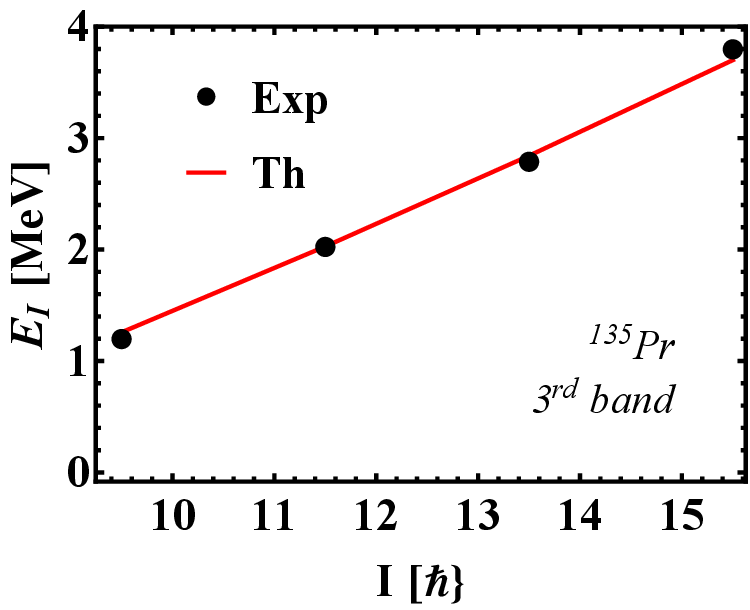}
\caption{ The excitation given by our calculations for the yrast band of $^{135}$Pr, using the MoI's, and $\theta$ determined by a fitting procedure.}
\end{center}
\label{Fig.8}
\end{figure}

From Table 9 we see that the MoI's ordering predicted by our calculations is: ${\cal J}_1>{\cal J}_3>{\cal J}_2$. However, due to the adopted fitting procedure this,  however, is a global result. In order to check whether this ordering holds also by using another fitting procedure, we fixed the MoI's by equating the calculated excitation energies for the lowest two states of band 1 and the second state of band 2, otherwise fixing 
$\theta$  to obtain a global best fit. In this way we found a set of MoI,s which reclaims a transverse wobbling regime for the odd system under consideration. Indeed, for $\theta =140^{0}$, the result is ${\cal J}_1=13.5285 [\hbar^2/MeV],\;{\cal J}_2=101.759 [\hbar^2/MeV],\;{\cal J}_3=52.9364 [\hbar^2/MeV]$. However, the overall agreement is of a poor quality. We also calculated the potential determined by the fitted parameters and I=19/2 which results in having the deepest minimum  at $q=0$, where $I_1=I$, which means that the angular momentum is oriented along the axis one. Therefore, ${\cal J}_2$= maximal, does not necessarily mean that the rotation axis is the 2-axis. The fact that the maximal MoI established by a global fit is ${\cal J}_1$,  indicates that for a larger angular momentum a change of MoI hierarchy takes place, and a new nuclear phase begins. 

The classical energy function exhibits several stationary points.The character of the stationary point to be minimum, maximum or saddle point is decided by the signs of the diagonal matrix elements of the Hessian: a) if all diagonal elements are positive, the stationary point is minimum; b) if all diagonal m.e. are negative, then we deal with a maximum, while; c) is a saddle point if one m.e. is positive and the other is negative.
Equating the Hessian to zero, one obtains the parameters $u$, and $v$ for which the critical points are degenerate. The resulting equations may be unified in a single formulae:
\begin{equation}
(1-u)(1-v^2)(v^2-u^2)(v^2-u)=0.
\label{separ}
\end{equation} 
The last factor in the above equation is obtained by equating the critical energies $E_{M}$, and $E_{s}$. Each factor generates a curve, called separatrices, in the parameter space spanned by (u,v).
As shown in Fig. 21, the separatrices are bordering  manifolds defining  unique nuclear phases characterized by a specific portrait of the stationary points. Indeed, among the factors involved in Eq.(\ref{separ}), we recognize those defining the two wobbling frequencies. On the other hand a vanishing energy defines a Goldstone mode \cite{Gold} which, as a matter of fact, render evidently a phase transition.
%\begin{center}

\begin{figure}[h!]
\begin{minipage}{7cm}
%\hspace*{1cm}
\includegraphics[height=4cm,width=0.7\textwidth]{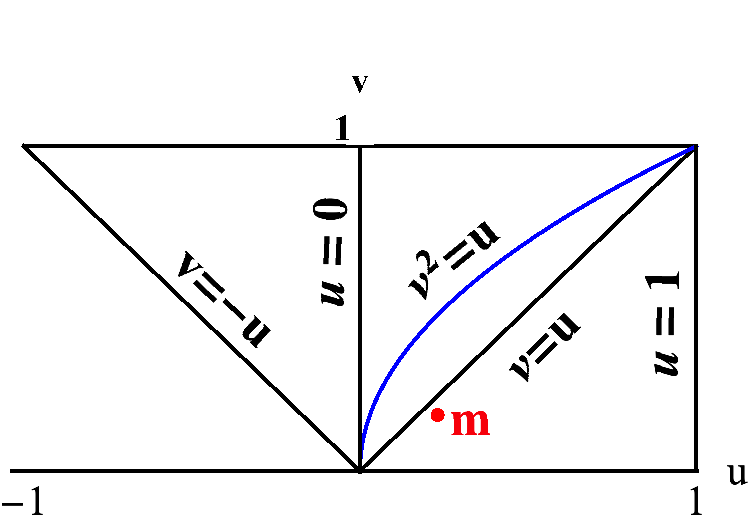}
\end{minipage}\ \ \hspace*{1cm}
\begin{minipage}{7cm}
\caption{ The phase diagram for I=15/2 of $^{135}$Pr. Using the MoI's, and $\theta$ determined by the adopted fitting procedure, and $\theta=150^{0}$, we mentioned the minimum point by a red and full circle having a lowercase $m$.}
\end{minipage}
\label{Fig.7}
\end{figure}
%\end{center}
\subsection{ Comment on the chiral features of the wobbling motion}
 In the angular momentum space, the change of sign of the a.m. defines a chiral transformation. A system is invariant to a chiral transformation if its rotational energy is preserved when the sense of rotation around an axis is changed.
Note that our starting Hamiltonian is a sum of two terms, one being symmetric to chiral transformations,$\hat{H}_s$ and one antisymmetric to chiral transformations, $\hat{H}_a$.

If $|\psi\rangle$ is an eigenstate for $\hat{H}_s$, and $C$ is a chiral transformation, then  $C|\psi\rangle$ is also eigenstate for $\hat{H}_{s}$, and corresponds to the same energy. In this case,
the function $|\psi\rangle$ has the chirality equal to one, since $C\psi\rangle=\psi\rangle$.
For $\hat{H}_a$, the above mentioned property changes to : If  $|\psi\rangle$ is an eigenstate of $\hat{H}_{a}$ corresponding to the eigenvalue $E$, then $C\psi\rangle$ is also eigenstate,
but corresponding to the energy $-E$. Therefore, the eigenvalues of  $\hat{H}_a$ split in two sets, one being the mirror image of the other one. This property is of a chiral nature. The eigenstates of $\hat{H}_a$ have the chirality -1 since $C|\psi\rangle=-|\psi\rangle$. The eigenstates of $\hat{H}_{rot}$ are mixtures of the two chiralities. When there are two sets of energies that are one the mirror image of the other, one says that a definite chirality is projected out \cite{Rad016}. In our calculation, the change of $\bf{I}\to-{\bf I}$ is achieved by changing $\theta$ to $\theta  +\pi$. The a.m. dependence of the wobbling frequencies corresponding to $\theta =-71^{0}$, and $\theta =109^{0}$ respectively, is shown in Fig. 22, left panel.
The two curves are parallel to each other, which suggests that the corresponding states have similar properties. The look of the potentials  $V$,  and $CVC^{-1}$, are shown in Figs.19, middle and right panels respectively, for $\theta=\pi/6$, and $\theta=7\pi/6$ respectively, and ${\cal J}_1:{\cal J}_2:{\cal J}_3=100:40:20 \hbar^2MeV^{-1}$. From these two potentials we may extract the symmetric, and antisymmetric parts of V.
\begin{equation}
V_s=\frac{1}{2}V(\pi/6)+V(7\pi/6);\;\;V_a=\frac{1}{2}V(\pi/6)-V(7\pi/6).
\end{equation}
The two potential of  definite chirality, are visualized in Fig. 22, middle panel and Fig. 22, right panel , respectively. A similar analysis can be performed also for the excitation energies. Indeed, Eq.(5.19) expresses explicitly
the dependence of the wobbling frequency on the angle $\theta$, which fixes the orientation of $\bf{j}$. Therefore, it is easy to calculate $\omega_I(\theta+\pi)$, with $\theta =-71^{0}$.
%%%%%%%%%%%%%%%%%%%%%%%%%%%%%%%%%%%%%%%%%%%%%%%%%%%%%%%%%%%%%%%%%%%%%%%%%%%%%%%%%%%%%%%%%%%%%%%%%%%%%%%%%%%%%%%%%%%%%%%%%%%%%%%
 The frequencies $\omega_I(\theta)$, and $\omega_I(\theta+\pi)$, with one being the chiral image of tThe phase diagramhe other one, are plotted in Fig. 22. for the yrast band. The two curves are parallel to each other, which suggests that the corresponding states have similar properties. However, they do not correspond to states of definite chirality. However, one can extract the symmetric, and antisymmetric terms of the excitation energy. Here we give the result for the yrast states:
{\scriptsize{
\begin{eqnarray}
&&E^{exc;1}_{I,s} = A_1 I^ 2 + \left(\omega_{I}(\theta)+ \omega_{I}(\theta+\pi)\right)/2-E_{11/2},\;E^{exc;1}_{I,a} = (2I+1)A_1 j_{1} -IA_2 j_2 + \left(\omega_{I}(\theta)-\omega_{I}(\theta+\pi)\right)/2,\nonumber\\
&&I=R+j,\;\;R=0,2,4,..,\theta=-71^{0}. 
\end{eqnarray}}} 
Since for asymmetric states, the energies $-E^{exc;1}_{I,a}$ are also eigenvalues of the antisymmetric Hamiltonian, we interpret the two sets of energies $E^{exc;1}_{I,s}$, and $-E^{exc;1}_{I,a}$ 
as defining two bands of chirality +1, and -1, respectively.  Although we don't have enough data to conclude that the  two bands are indeed of real chiral type, however, due to the above mentioned features, they might be considered as germinos of chiral bands.
Indeed, there are properties unanimously accepted, which prevent us to make a decisive statement on this matter. For a chiral band the system rotates around an axis, which doesn't belong to any of principal planes, while here the rotation axis is a principal axis. However, since in our case the Hamiltonian involves linear terms in the total angular momentum, the wobbling motion, and chiral properties seem not  to be disconnected.

\begin{figure}[h!]
%\hspace*{-2cm}
\begin{center}
\includegraphics[height=4cm,width=0.4\textwidth]{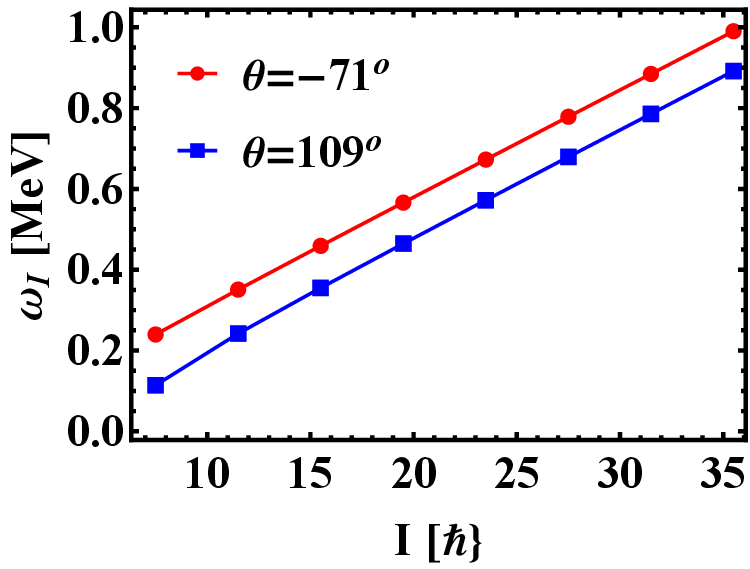}\includegraphics[height=4cm,width=0.4\textwidth]{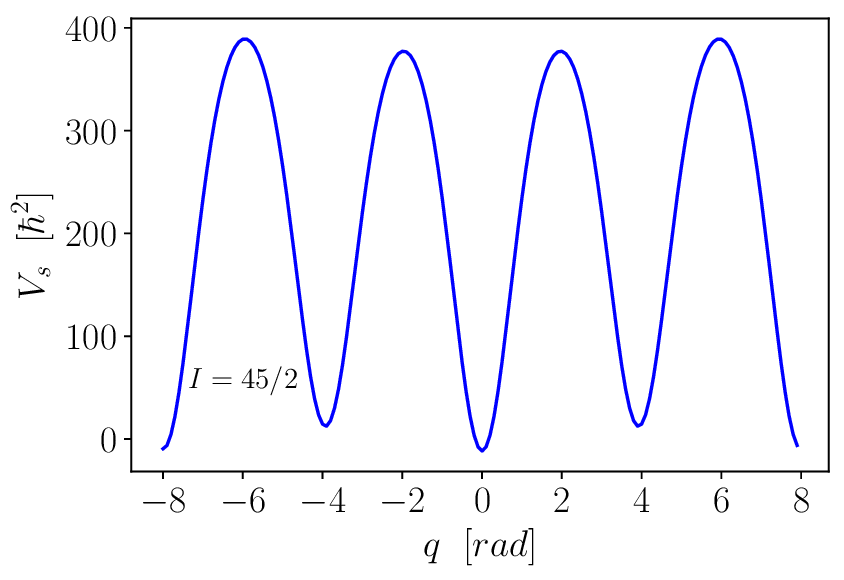}\includegraphics[height=4cm,width=0.4\textwidth]{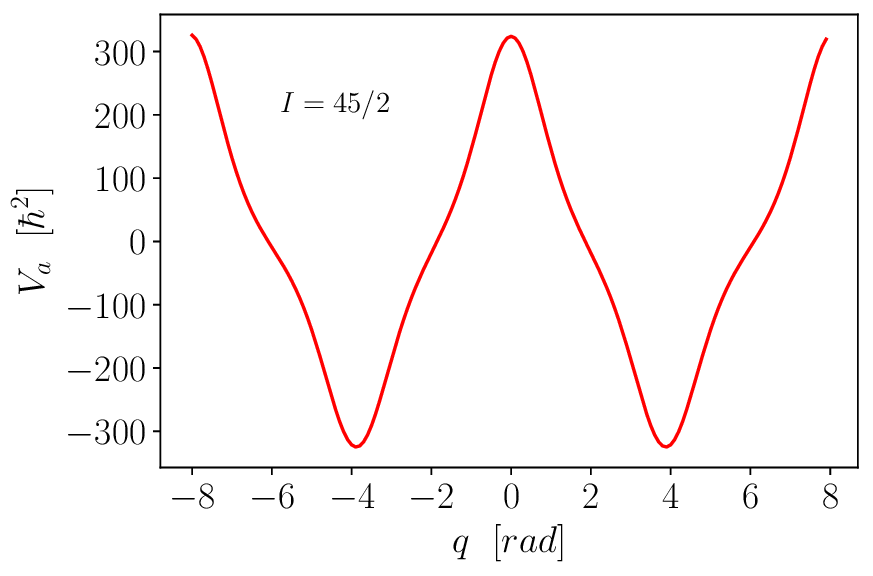}
\caption{ The wobbling frequencies corresponding to the angles $\theta=109^0$, and $\theta=-71^0$, respectively{left panel}. The symmetric potential, with respect to the chiral transformations, as function of q (middle panel). The antisymmetric potential, with respect to the chiral transformations, as function of q (right panel).
} 
\end{center}
\label{Fig.11}
\end{figure}

Within this formalism we described  the experimental data for the electric quadrupole intra- and inter-band transitions as well as the magnetic dipole transitions. As for the E2 transitions , we need the wave functions describing the involved states, and the transition quadrupole operator. The wave function for an I-state is the solution of the 
Scr\"{o}dinger equation for the given total angular momentum, I. Note that the wave function is degenerate with respect to "M", the projection of I on the x-axis, in the laboratory frame. Since the ground state is the vacuum state for the wobbling phonon operator, and moreover, in the minimum point of the constant energy surface, the a.m. projection on the one-axis of the intrinsic frame is equal to -I, it results that the K quantum number is equal to -I. Therefore the solution of the Schr\"{o}dinger equation must be labelled by  the mentioned  quantum numbers, i.e. 
\begin{equation}
\Psi_{IM}=\Phi_{I,-I}|IM,-I\rangle, \;\rm{with}\;\;|IMK\rangle =\sqrt{\frac{2I+1}{8\pi^2}}D^I_{M,K}.
\end{equation}
Here we consider the first two wobbling bands as signature partner bands, the arguments being in detail given in Ref.\cite{PoRa}. More specifically, the spin sequence of the first band is 
$j+R$, for R=0,2,4,... , while for the second band the spin succession is $j+R$, with R=1,3,5,....  Note that the quadrupole inter-band transition is forbidden since, for the states mentioned above, have $\Delta K=1$. In this case, considering the component $K=-I+1$ in one of the involved states is necessary. The quadrupole transition operator is taken as:
\begin{equation}
{\cal M}(E2;\mu)=\sqrt{\frac{5}{16\pi}} e\left(Q_0D^2_{\mu 0}+Q_2\left(D^2_{\mu 2}+D^2_{\mu -2}\right)\right),
\end{equation}The phase diagram
where $Q_0$, and $Q_2$, denote the $K=0$, and $K=\pm 2$ components of the quadrupole transition operator, respectively.  Note that since the intrinsic component of the wave function depends on one of the conjugate variables $q$, and $d/dq$, that is $q$, we must express the quadrupole operators in terms of the q variable. This will be achieved by writing Q-s in the space of angular momentum and then use the Bargmann representation of the a.m. components. Thus we have:
\begin{equation}
Q_0=-\frac{1}{4}\sqrt{\frac{2}{3}}\left(\hat{I}_{+}\hat{I}_{-}+\hat{I}_-\hat{I}_+\right)+\sqrt{\frac{2}{3}}\hat{I}_1^2,\;\;Q_2=\frac{1}{2}\left(\hat{I}_{+}^{2}+\hat{I}_{-}^{2}\right).
\label{qtran}
\end{equation}
The reduced E2 transition probability is expressed as the square of the reduced matrix element of the quadrupole transition operator.
The magnetic dipole transition operator is:
\begin{eqnarray}
{\cal M}(M1;\mu)=\sqrt{\frac{3}{4\pi}}\mu_N\sum_{\nu}\left(g_R\hat{R}_{\nu}+g_j\hat{j}_{\nu}\right)D^{1}_{\mu \nu}\equiv M^{coll}_{1\mu}+M^{sp}_{1\mu},
\end{eqnarray}
where $R_{\nu}$, and $j_{\nu}$ are the spherical components of the core and the odd nucleon angular momenta, respectively. $g_R$ and $g_j$ stand for the gyromagnetic factors of the core, and the coupled odd nucleon, respectively. Also, the standard notation for the Wigner function, $D^{J}_{MK}$, and for the nuclear magneton, $\mu_{N}$, are used.
To calculate the collective part of the transition matrix element, we need to express the wave function describing the odd system as a Kronecker product of the core, and the odd particle wave functions:
\begin{equation}
|IMK\rangle =\frac{1}{2j+1}\sum_{M_R,\Omega,R} C^{R\;j\;I}_{M_R\;\Omega\;M }|RM_RK\rangle \psi_{j\Omega}.
\end{equation}

To calculate the reduced m.e. of the single particle M1 operator we need the wave function describing the odd proton whose a.m. is placed in the plane XOY making the angle $\theta$ with the axis OX. This function is obtained by rotating around the axis 3, the function $\psi_{j,j}$  associated to the odd proton having the a.m. along the 1-axis,i.e.,$\psi_{j}^{\prime}=R_3(\theta)\psi_{jj}.$
Finally, the magnetic dipole reduced transition probability is given by:
\begin{equation}
B(M1;I\to I')=\left[\langle \Psi_{I}||{\cal M}(M1)||\Psi_{I'}\rangle\right]^2.
\end{equation}
Results for the branching ratios of E2 and M1 transitions are collected in Table 10.
\begin{center}
\begin{table}[h!]
\begin{tabular}{|c|cc|cc|cc|}
\hline
&\multicolumn{2}{c|}{$\frac{B(E2;I^-\to (I-1)^-)}{B(E2;I^-\to (I-2)^-)}$} &\multicolumn{2}{c|}{$\frac{B(M1;I^-\to (I-1)^-)}{B(E2;I^-\to (I-2)^-)}[\frac{\mu_N^2}{e^2b^2}]$}& \multicolumn{2}{c|}{$\delta_{I^-\to (I-1)^-}$ [MeV.fm]}\\
\hline
$I^{\pi}$        & Exp.           &    Th.                                &    Exp.     &Th.                                                       &Exp.     &     Th.  \\
\hline
$\frac{21}{2}^{-}$&0.843$\pm$0.032    &0.510&0.164$\pm$ 0.014&0.164&-1.54$\pm$ 0.09& -0.542\\
$\frac{25}{2}^{-}$&0.500$\pm$0.025    &0.500&0.035$\pm$ 0.009&0.066&-2.384$\pm$0.37& -0.703\\
$\frac{29}{2}^{-}$&$\ge$0.261$\pm$0.014&0.487&$\le$ 0.016$\pm $0.004&0.033&-        & -0.873\\
$\frac{33}{2}^{-}$&-                &0.473&-                 &0.019&-        & -1.052\\
\hline
\end{tabular}
\caption{\textmd{The calculated branching ratios $B(E2)_{out}/B(E2)_{in}$, and $B(M1)_{out}/B(E2)_{in}$ as well as the mixing ratios $\delta$ are compared with the corresponding experimental data taken from Ref. \cite{Matta}.}}
\label{Table II}
\end{table}
\end{center}
\section{Summary and Conclusions}
Here we presented few selected issues of our contribution to the field of wobbling and chiral motion in atomic nuclei. Since one deals with high spin state we chose a semi-classical approach.
Moreover the trial function involved in the variational equation is a coherent state for SU(2) group generated by the angular momenta determining a rigid rotor Hamiltonian, for the even-even system  and a product of SU(2) coherent states the factors being associated to the core and the odd nucleon for an even-odd system.For the even-even system a new quantization method for the classical orbitals is proposed, which results in a Sch\"{o}dinger equation with fully separable kinetic and potential energies terms. In the harmonic approximation for the potential energy term, a compact formulae for the wobbling frequency, different from those proposed by other authors is obtained. An excellent agreement with experimental data is obtained for both excitation energies and B(E2) values measured for $^{158}$Er. 

As for the even-odd nuclei to wobbling frequencies are obtained which reflects the fact that the angular momentum of the odd nucleon is frozen. The formalism is was applied to $^{163}$Lu where plenty of experimental data are available. Two scenarios of defining the bands TSD1, TSD2, TSD3 and TSD4 are proposed. In one case each state member of the four bands is treated variatinally with j=13/2 and the core states of positive parity for the first three bands and negative parity for TSD4. In the other scenarios the first three bands are defined in a similar way but the odd nucleon for TSD4, moves in the orbital j=$h_{9/2}$. Both variants yield results for excitation energies, transition probabilities, mixing ratios, branching ratios, angular momentum alignment, excitation energies relative to an rigid rotor effective energy, dynamic moment ofThe phase diagram inertia  which are in good agreement with the corresponding data. The classical orbital dependence on energy and angular momentum is studied by intersecting the surfaces of constant energy and constant angular momentum, respectively.The contour plots for all the four bands as well as the phase diagram are given. 

A simultaneous description of wobbling and chiral motion is presented in Section IV.
If in the particle-triaxial rigid rotor Hamiltonian the coupling potential is neglected, due to the freezing of the particle angular momentum, one arrives to a cranked rotor Hamiltonian. The linearized equations of motion for the cranked Hamiltonian admit solutions which describe an a-planar motion. One derives analytical expressions for the three wobbling frequencies, for both the cranked and chirally transformed Hamiltonians. Then one builds up two chiral bands whose properties are analyzed by means of the corresponding contour plots. Thu one proves that the two signatures, wobbling and chiral, may coexist.

In the last section, the odd system is treated by using a new boson expansion for the total angular momentum components while the angular momentum of the odd nucleon is frozen in a principal plane. The maximal moment of inertia  corresponds to the axis 2. The classical trajectories are quantized by writing the boson expanded Hamiltonian in the Bargmann representation.
The resulting Schr\"{o}dinger Hamiltonian comprises a potential having a complex structure exhibiting several symmetric deepest minima and local minima. It seems that the local minimum is a chiral transformation image of the deepest minimum. The harmonic approximation is used to interpret the experimental data concerning the excitation energies and the transition probabilities in three bands of $^{135}$Pr.A comment about the transition fro the traversal to the longitudinal regime, is included. 

The results of this chapter recommend the semi-classical formalism as an useful tool for describing the main features of wobbling and chiral motion, in a realistic fashion. 

\end{document}